\documentclass[aps.pra,onecolumn,showpacs,preprintnumbers,amsmath,amssymb]{revtex4}
\usepackage{mathrsfs}
\usepackage{graphicx}
\usepackage{amsmath}
\usepackage{ulem}
\usepackage{color}

\begin{document}

\title{Exact solution for the inhomogeneous Dicke model in the canonical ensemble: thermodynamical limit and finite-size corrections}
\author{W. V. Pogosov$^{1,2,3}$,  D. S. Shapiro$^{1,3,4,5}$, L. V. Bork$^{1,6}$, A. I. Onishchenko$^{7,3,8}$}
\affiliation{$^1$N. L. Dukhov All-Russia Research Institute of Automatics, Moscow, Russia}
\affiliation{$^2$Institute for Theoretical and Applied Electrodynamics, Russian Academy of Sciences, Moscow, Russia}
\affiliation{$^3$Moscow Institute of Physics and Technology, Dolgoprudny, Russia}
\affiliation{$^4$V. A. Kotel'nikov Institute of Radio Engineering and Electronics, Russian Academy of Sciences, Moscow, Russia}
\affiliation{$^5$National University of Science and Technology MISIS, Moscow, Russia}
\affiliation{$^6$Institute for Theoretical and Experimental Physics, Moscow, Russia}
\affiliation{$^7$Bogoliubov Laboratory of Theoretical Physics, Joint Institute for Nuclear Research, Dubna, Russia}
\affiliation{$^8$Skobeltsyn Institute of Nuclear Physics, Moscow State University, Moscow, Russia}

\begin{abstract}
We consider an exactly solvable inhomogeneous Dicke model which describes an interaction between
a disordered ensemble of two-level systems with single mode boson field. The existing method
for evaluation of Richardson-Gaudin equations in the thermodynamical limit is extended to the case of Bethe equations in Dicke model. Using this extension, we present expressions both for the ground state and lowest excited states energies as well as leading-order finite-size corrections to these quantities for an arbitrary distribution of individual spin energies. We then evaluate these quantities for an equally-spaced distribution (constant density of states). In particular, we study evolution of the spectral gap and other related quantities. We also reveal regions on the phase diagram, where finite-size corrections are of particular importance.
\end{abstract}

\pacs{02.30Ik, 42.50.Ct, 03.65.Fd}

\author{}
\maketitle
\date{\today }

\section{Introduction}

Recent progress in engineering of artificial quantum systems for information technologies renewed an interest to Dicke (Tavis-Cummings) model \cite{Dicke} and its exact solution,
already well known for a long time \cite{Gaudin, Lieb, Wiegmann, Yudson, Dukelsky, Skrypnyk}. Dicke model describes an interaction between a collection of two-level systems and single mode radiation field, while physical realizations range from superconducting qubits coupled to microwave resonators to polaritons in quantum wells, see e.g. Refs. \cite{Balili, Zagoskin} and references therein; furthermore, it can be also applied to Fermi-Bose condensates near the Feshbach resonance \cite{Andreev}.

The characteristic feature of macroscopic artificial quantum systems such as superconducting qubits is a disorder in excitation frequencies and inhomogeneous broadening of the density of states. This feature is due to fundamental mechanisms: for example,
an excitation energy of flux qubits depends exponentially on Josephson energies \cite{qubit},
which makes it extremely sensitive to characteristics of nanometer-scale Josephson junctions.
Inhomogeneous broadening appears even in the case of microscopic two-level systems,
such as NV-centers, where it is induced by spatial fluctuations of background magnetic moments \cite{Lukin}. In the case of NV-centers, the density of states is characterized by the $q$-Gaussian distribution \cite{Krimer}.
Moreover, there are prospect to utilize the broadening for the construction of a multimodal quantum memory \cite{Molmer}. It is also possible to engineer a density of states profile by using, e.g., a so-called spectral hole burning technique which allows to perform a significant optimization of various characteristics of spin-photon hybrid systems \cite{Krimer}.

Inhomogeneous Dicke model, which explicitly takes into account a disorder in excitation energies, has been studied in Ref. \cite{Littlewood} using a mean-field treatment within functional-integral representation of the partition function. This study revealed an existence of a rather rich phase diagram. The interaction between boson and spin subsystems gives rise to a finite gap in the energy spectrum between the first excited state and the ground state. It has an apparent similarity with the superconducting gap in the Bardeen-Cooper-Schrieffer (BCS) theory of superconductivity. The mean-field approximation for the inhomogeneous Dicke model also becomes exact in the thermodynamical limit, as for the BCS pairing Hamiltonian. However, this approximation is expected to fail in the mesoscopic regime, which seems to be more relevant for near-future technological applications with macroscopic artificial 'atoms', such as superconducting qubits. Indeed, structures, which consist of tens or hundreds of superconducting qubits and show signatures of global coherence, have been successfully fabricated and explored very recently \cite{Macha,Zagoskin}. Such structures are often refereed to as superconducting metamaterials.

The mesoscopic regime of the Dicke model as well as the emergence of the macroscopic limit can be properly described only using approaches based on a canonical ensemble, which takes into account that the 'particle' number is fixed. This circumstance makes it difficult to apply standard mean-field methods. By 'particle' number one should understand the total number of bosons and excited two-level systems (spins), so they can be referred to as pseudo-particles. The limiting validity of grand canonical description is well known in the case of pairing correlations in ultrasmall metallic systems at low temperatures and in nuclei, for which usual mean-field approximation can give results inadequate even on a qualitative level \cite{Ralph}. For example, it predicts vanishing of superconducting correlations below certain mean interlevel distance, while more advanced approaches show that they do not disappear. One of such approaches is to turn to Richardson-Gaudin solution of BCS pairing Hamiltonian \cite{Rich1,Gaudin} via Bethe ansatz technique, which was utilized to evaluate various characteristics along the crossover from few-particle systems to the macroscopic regime, see, e.g., Refs. \cite{Schechter,Duk}. In particular, finite-size corrections can be elaborated iteratively using the electrostatic analogy for Bethe equations in the thermodynamical limit \cite{Rich3,Altshuler}. For an example of a recent application of this exact solution for the evaluation of form factors, see Ref. \cite{Bet}, while the extension of this approach to other pairing models was reported in Ref. \cite{Amico}.

Dicke model belongs to the same class of exactly solvable quantum models as Richardson and Gaudin models \cite{Gaudin, Talalaev, Loss, Ghent}. To a certain extent, it can be viewed as Richardson model in which interaction between spins is mediated by a bosonic degree of freedom. However, in contrast to the Richardson model, Dicke model supports arbitrarily large number of pseudo-particles through this degree of freedom.
This fact sometimes makes it not so straightforward to apply ideas relevant for Richardson-Gaudin models
to the Dicke model, see, e.g., recent developments on the particle-hole duality \cite{Pogosov,Faribault,Bork}.
 The phase diagram of the inhomogeneous Dicke model in the thermodynamical limit is much richer, since it contains larger number of controlling parameters which include pseudo-particle density, mean detuning between the spin and boson energies, as well as spin-boson coupling energy \cite{Littlewood}.

The aim of the present paper is to construct a solution of Bethe equations for inhomogeneous Dicke model in the thermodynamical limit in the spirit of the approach developed by Richardson, as well as to evaluate, within the canonical ensemble, leading finite-size corrections both to the ground state energy and low-energy excited states. Although we basically follow the approach of Richardson \cite{Rich3}, many aspects of the derivation are different, since Bethe equations for Dicke model include a term of a new (divergent) type, which significantly modifies the multipole expansion encoded into the approach and also alters a nonlinear equation for the electrostatic field.

We then apply the derived formulas for the simplest case of an equally-spaced distribution of spin energies. We reveal an existence of the rich phase diagram and find regions of parameters, where finite-size corrections are of a particular importance.

\section{Preliminaries}

We consider a Hamiltonian of the form
\begin{eqnarray}
H=\sum_{n}^L 2 \epsilon_n s_{n}^z + \omega b^{\dagger } b + g \sum_{n}^L (b^{\dagger }s_{n}^- + b s_{n}^+),
\label{Hamiltonian}
\end{eqnarray}
where $b^{\dagger }$ and $b$ correspond to the boson degree of freedom:
\begin{eqnarray}
[b, b^{\dagger }]=1,
\label{boson}
\end{eqnarray}
while $s_{n}^z$ and $s_{n}^\pm$ correspond to the paulion degrees of freedom and describe a set of $L$ two-level systems:
\begin{eqnarray}
& [s_{n}^+, s_{n}^-]=2s_{n}^z, \\&
[s_{n}^z, s_{n}^\pm] = \pm s_{n}^\pm.
\label{paulion}
\end{eqnarray}

This Hamiltonian commutes with the operator of the total pseudo-particle number, i.e., the number of bosons plus the number of excited two-level systems. Let us denote this number as $M$. We then consider a limit $L \rightarrow \infty$, while $M$ behaves in the same way, so that $M/L$ is constant. Let us also assume that the density of energies $\epsilon_n$ grows at $L \rightarrow \infty$, whereas $\omega$ is independent on $L$ and $g$ scales as $1/\sqrt{L}$. For any fixed $M$, there are different eigenstates of the Hamiltonian. The lowest energy state is the ground state at given $M$, while others represent excited states.

Bethe equation for each rapidity $e(i)$ ($i$ ranges from 1 to $M$) reads \cite{Gaudin, Talalaev, Loss, Ghent, Gritsev}
\begin{eqnarray}
\frac{2e(i)}{g^2}-\frac{\omega}{g^2}+\sum_{j\neq i}^M \frac{1}{e(i) - e(j)} - \frac{1}{2}\sum_{n}^L \frac{1}{e(i) - \epsilon_n}=0,
\label{Bethe}
\end{eqnarray}
while the energy of the system is expressed through the roots $e(i)$ as
\begin{eqnarray}
E = 2 \sum_{i}^M e(i).
\label{energybethe}
\end{eqnarray}
The set of equations (\ref{Bethe}) differs from Richardson equations by presence of the terms $2e(i) / g^2$. Hereafter we use notations similar to the ones of Ref. \cite{Rich3} in order to facilitate a comparison. Notice that within our notations qubit excitation energy is $2\epsilon_n$.

\section{Electrostatic analogy}

There exists a two-dimensional electrostatic analogy for the set of Bethe equations (\ref{Bethe}), which may be treated as equilibrium conditions for the Coulomb plasma \cite{Gaudin, Rich3}. Roots $e(i)$ can be interpreted as locations of $M$ free charges of unit strength in the complex plane. They repeal each other, but they are also attracted by $L$ fixed charges of $-1/2$ strength positioned at $\epsilon_n$ on real axis. In addition, free charges are subjected into two forces produced by the uniform external field $-\omega / g^2$ and by parabolic confining potential $2e(i)/g^2$. The last term is absent in both Richardson and Gaudin models. It modifies significantly the approach developed by Richardson for the thermodynamical limit \cite{Rich3}.

Note that the analogy with the Coulomb plasma can be used to construct a kind of a probabilistic approach to the solution of Richardson-Gaudin equation having connections with a conformal field theory \cite{Pogosov}, but we are not going to pursue this issue here.

Let us consider a function
\begin{eqnarray}
F(z)=\sum_{j} \frac{1}{z - e(j)}-\frac{1}{2}\sum_{n} \frac{1}{z - \epsilon_n}
-\frac{\omega}{g^2}+\frac{2z}{g^2},
\label{Fz}
\end{eqnarray}
which represents an electrostatic field of the whole system of charges. It includes a term $2z/g^2$ absent in the case of Richardson model.

Using (\ref{Bethe}), it is not difficult to find that $F$ satisfies the equation
\begin{eqnarray}
F^2 + \frac{dF}{dz}=\frac{2}{g^2}+\frac{1}{2}\sum_{n} \frac{1}{(z - \epsilon_n)^2}+
\left( \frac{1}{2}\sum_{n} \frac{1}{z - \epsilon_n}+\frac{\omega}{g^2}-\frac{2z}{g^2}\right)^2+
\frac{4M}{g^2}-\sum_{n} \frac{H(n)}{z - \epsilon_n}.
\label{Rikatya}
\end{eqnarray}
where
\begin{eqnarray}
H(n)=- \sum_{j} \frac{1}{e(j) - \epsilon_n}.
\label{Hn}
\end{eqnarray}
It can be rewritten as
\begin{eqnarray}
H(n)=-\frac{1}{2\pi i }\oint_{C_{e}}  \frac{F(z)dz}{z - \epsilon_n},
\label{Hnint}
\end{eqnarray}
where a closed contour $C_{e}$ encloses all the singularities of $F$ coming from poles at all roots $e(j)$ and excludes those poles, which are due to the set of energies $\epsilon_k$.

In Appendix A, we develop two expansions of the field $F(z)$, which are multipole expansion and series expansion in powers of $1/L$. They are further used to derive leading-order solution as well as finite-size corrections.

\section{General solution}

It is further assumed that free charges in $L\rightarrow \infty$ limit merge into a line of charge, so that the poles of $F$ due to the first term in (\ref{Fz}) form a branch cut, which extends from point $a=\lambda + i \Delta$ to $a^{*}=\lambda - i \Delta$ in the complex plane. Let us consider the following ansatz for the leading-order in $1/L$ contribution $F_{0}$ to the total field $F(z)$
\begin{eqnarray}
F_{0} = - \frac{Z}{2} \left( \sum_{n} \frac{1}{\eta _n (z - \epsilon_n)}-\frac{1}{\xi^2}
\right ),
\label{F0ansatz}
\end{eqnarray}
where
\begin{eqnarray}
& Z = \sqrt{(z-a)(z-a^{*})},
\notag \\&
\eta _n= \sqrt{(\varepsilon_n-a)(\varepsilon_n-a^{*})}=\sqrt{(\varepsilon_n-\lambda)^2+\Delta^2}.
\label{Zetaabbr}
\end{eqnarray}
This ansatz readily follows once we take into account that $F_0$ has a single cut on an interval $[a, a^{*}]$ as well as poles with known residues at $\epsilon_n$ and infinity. It differs from the similar ansatz in Richardson model by the last term in the right-hand-side of (\ref{F0ansatz}), which is dictated by the presence of nonzero moment $F^{(-1)}$.

By substituting (\ref{F0ansatz}) to the field equation (\ref{RikatyaF0}) in leading order and using a mathematical trick outlined in Appendix B, we find that (\ref{F0ansatz}) is indeed a solution provided three relations are satisfied:
\begin{eqnarray}
& \frac{1}{\xi^2} = \frac{4}{g^2}, \label{ksai}\\&
\sum_{n}
\frac{1}{\eta _n}=\frac{2}{g^2}(\omega-2 \lambda), \label{gap}\\&
\sum_{n} \frac{\epsilon_n-\lambda}{\eta _n}=\frac{2 \Delta ^2}{g^2}+(L-2M).
\label{gapchemeqs}
\end{eqnarray}
Last two equations are actually gap equation (or the equation for the order parameter) and the equation for the chemical potential, which were obtained before in the thermodynamical limit in Ref. \cite{Littlewood} using mean-field method. These two equations must be solved together. They have similarities with the corresponding equations for the BCS theory, but the interaction constant in the gap equation (\ref{gap}) is now dependent on the chemical potential. In its turn, the left-hand-side of the equation for the chemical potential (\ref{gapchemeqs}) contains a contribution $2 \Delta ^2 / g^2$, which is a mean number of bosons in the ground state \cite{Littlewood}. The relation between the solution of Eqs. (\ref{gap}), (\ref{gapchemeqs}) and spectral gap is clarified below.

We would like to stress that the gap equation and the equation for the chemical potential appear in other powers of the expansion of both sides of the field equation over powers of $z$, as compared to Richardson model \cite{Rich3}. This is due to the presence of the term $\sim z$ in the expression for the field $F$ and the nonlinearity of the field equation itself.

Using the approach of Appendix B and the result of a multipole expansion (\ref{EjF2}), we find the ground state energy in the leading order
\begin{eqnarray}
E_{gr0}=\sum_{n} (\epsilon_n - \eta _n) +
\lambda (2M-L)+\frac{\Delta^2}{g^2}(\omega-2\lambda).
\label{energygr0}
\end{eqnarray}
A first order in $1/L$ correction to this quantity is obtained in Appendix C using the expansion of the field equation. It can be cast into a compact form as
\begin{eqnarray}
E_{gr1}=
\sum_{l} \zeta_l - \sum_{n}  \eta_n + \frac{1}{2} (2\lambda-\omega),
\label{sumh1final}
\end{eqnarray}
where $\zeta_l=Z(x_l)$ and $x_l$ are real zeros of $F_0$.

Within the electrostatic mapping, excited states with one excitation correspond to a single isolated charge positioned out of the line of charges. A detailed analysis of this situation is presented in Appendix D. This single charge only slightly disturbs a whole configuration of charges. In leading order in $1/L$, its allowed positions are given by real zeros $x_l$ of $F_0$. This requirement has a simple meaning that the force acting on the isolated charge from the line of charges must be zero. As it is shown in Appendix D, the isolated charge provides a contribution to the total energy given by $2\zeta_l=2\sqrt{(x_l-\lambda)^2 + \Delta^2}$. This is nothing but the excitation energy in leading order.

It is of importance that zeros $x_l$ are confined between two neighboring energies $\epsilon$, as follows from $F_0(z)=0$. Consequently, in the thermodynamical limit, $x_l$ can be replaced by $\epsilon$ due to the infinite density of these energy levels. However, in the case of a density of states profile, having holes or abrupt terminations, this is not always true. Such situations must be analyzed with a special care. In the next Section, we consider a system with the constant density of states within a finite extension, which represents an example of such a distribution.

The above result for the excitation energy is similar to the well known result of Bardeen-Cooper-Schrieffer theory of superconductivity. Note that in this theory chemical potential resides exactly in the middle of the interaction band, so that the equation for the chemical potential is fulfilled automatically and therefore is dropped; however, it must be kept in the situation of a crossover from local Bose-condensed pairs to the dense condensate \cite{Eagles, Leggett,Pogosov,Bork,We}.

We would like also to stress that the obtained result for the excitation spectrum is quite different from the one for the Dicke model without inhomogeneous broadening, see, e.g., Ref. \cite{Garraway}. In the case of homogeneous model, the eigenenergies for a given $M$ are located nearly equidistantly with the separation $\gtrsim g \sqrt{M} \sim L^0$ between the two neighboring values. For the inhomogeneous model, finite separation (gap) generally can survive only for the energy difference between the first excited and ground states.

The evaluation of leading finite-size correction to the excitation energy $\Delta_1$ is the same as for the Richardson model \cite{Rich3}. The result is also identical except of the fact that all input functions must be modified. We, therefore, present these results in Appendix D without a derivation.

The obtained expressions for the ground state and excited states energies are completely generic, so that various distributions of spin energies $\epsilon$ can be used to evaluate them. Among physically meaningful distributions of the density of states are the Gaussian and Lorentzian  distributions \cite{Littlewood} or $q$-Gaussian distribution relevant for NV-centers \cite{Krimer}. In Ref. \cite{Littlewood} such smooth distributions were supplemented by a lower cut-off in order to avoid some nonphysical effects. The simplified equally-spaced distribution of $\epsilon$ between the two cut-offs would be also of interest. Physically it might correspond to the broad distribution, for which only central part, where the density of states is nearly constant, is left, while the remaining part is 'burnt'. Alternatively, for artificial macroscopic 'atoms' such as superconducting qubits, it can be achieved by a proper fabrication/selection of these 'atoms'. Anyway, equally-spaced distribution is a good starting point to analyze finite-size corrections. It is also useful in the view of establishing of a connection with the problem of pairing correlations in superconductors. The calculations for such a distribution for the case of Richardson model and at fixed $M/L=1/2$ were performed in Ref.  \cite{Altshuler}. Such calculations for the Dicke model are presented in the next Section for a general $M/L$.

\section{Equally-spaced distribution}

In this section, we analyze the case of equally-spaced distribution of spin energies $\epsilon_n$. We assume that these energies are confined between two cutoffs $E_1$ and $E_2 = E_1 + \Omega$; hence, $\epsilon_n = E_1 + n d$, where $d=\Omega / L$.

\subsection{Order parameter and chemical potential}

Let us consider solutions of the equations for the order parameter (\ref{gap}) and chemical potential (\ref{gapchemeqs}). In the thermodynamical limit, sums in (\ref{gap}) and (\ref{gapchemeqs}) can be approximated by integrals. We obtain within this approximation
\begin{eqnarray}
& \log \frac{\frac{\Omega'}{2}-\lambda'+\sqrt{(\frac{\Omega'}{2}-\lambda')^2+\Delta'^2}}
{\frac{-\Omega'}{2}-\lambda'+\sqrt{(\frac{\Omega'}{2}+\lambda')^2+\Delta'^2}} = 2 \Omega' (\omega'-2\lambda') \label{gaplog} \\&
\sqrt{(\frac{\Omega'}{2}-\lambda')^2+\Delta'^2} - \sqrt{(\frac{\Omega'}{2}+\lambda')^2+\Delta'^2}
=2 \Omega' \left(\Delta'^2+(\frac{1}{2}-\frac{M}{L})\right),
\label{chemelog}
\end{eqnarray}
where $\Delta' = \Delta/g\sqrt{L}$, $\lambda'=(\lambda - (E_1+E_2)/2)/g\sqrt{L}$,
and $\Omega'= \Omega/g\sqrt{L}$, $\omega'=(\omega - (E_1+E_2))/g\sqrt{L}$ are the dimensionless gap,
chemical potential, width of spin energies distribution, and the detuning, respectively.
Equations (\ref{gaplog}) and (\ref{chemelog}) are transcendental equations which cannot be solved explicitly
in contrast to similar equations for BCS pairing model. However, the solution can be readily obtained numerically.

\begin{figure}[h]
\includegraphics[width=0.45\linewidth]{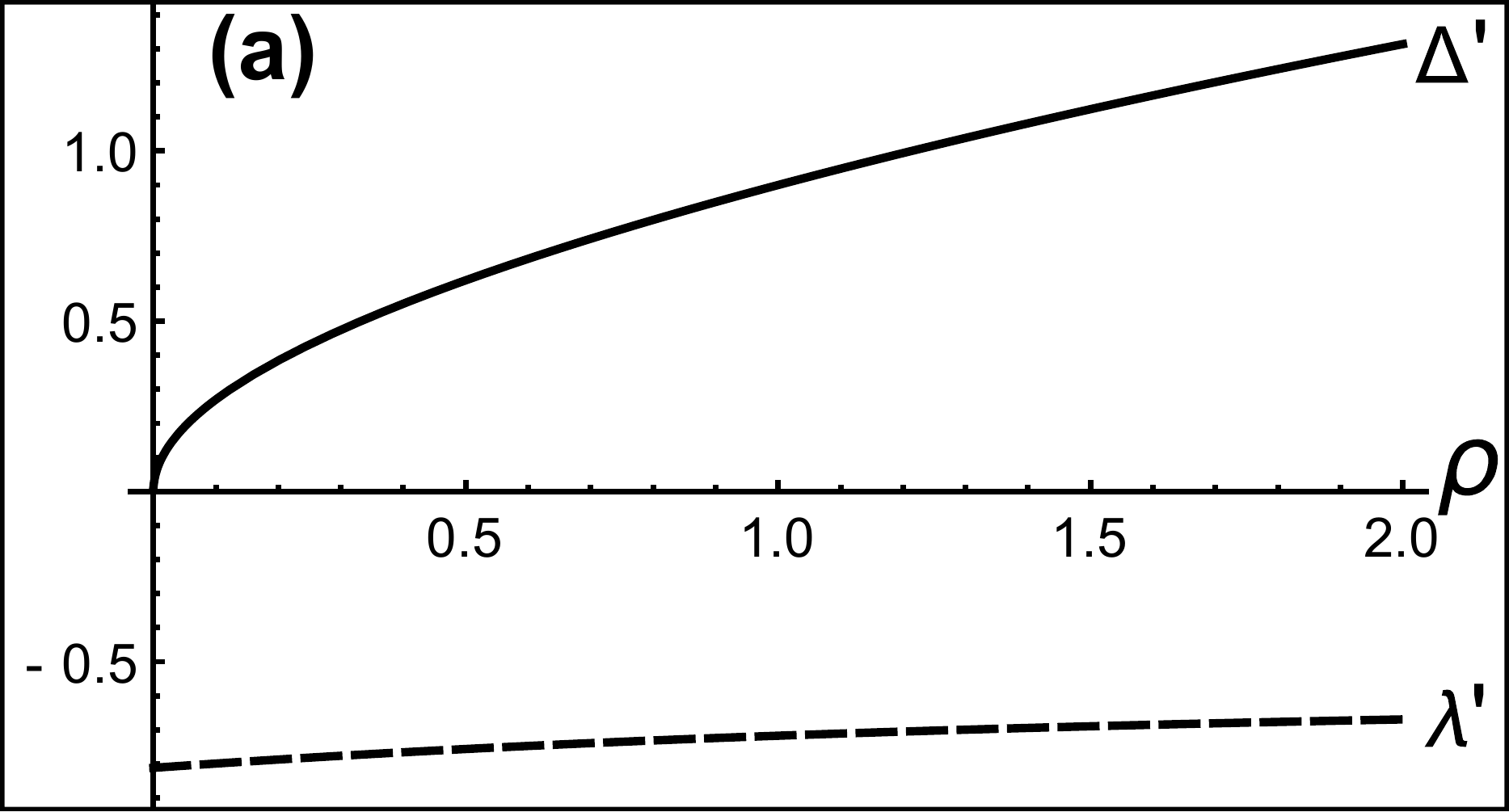}
\includegraphics[width=0.45\linewidth]{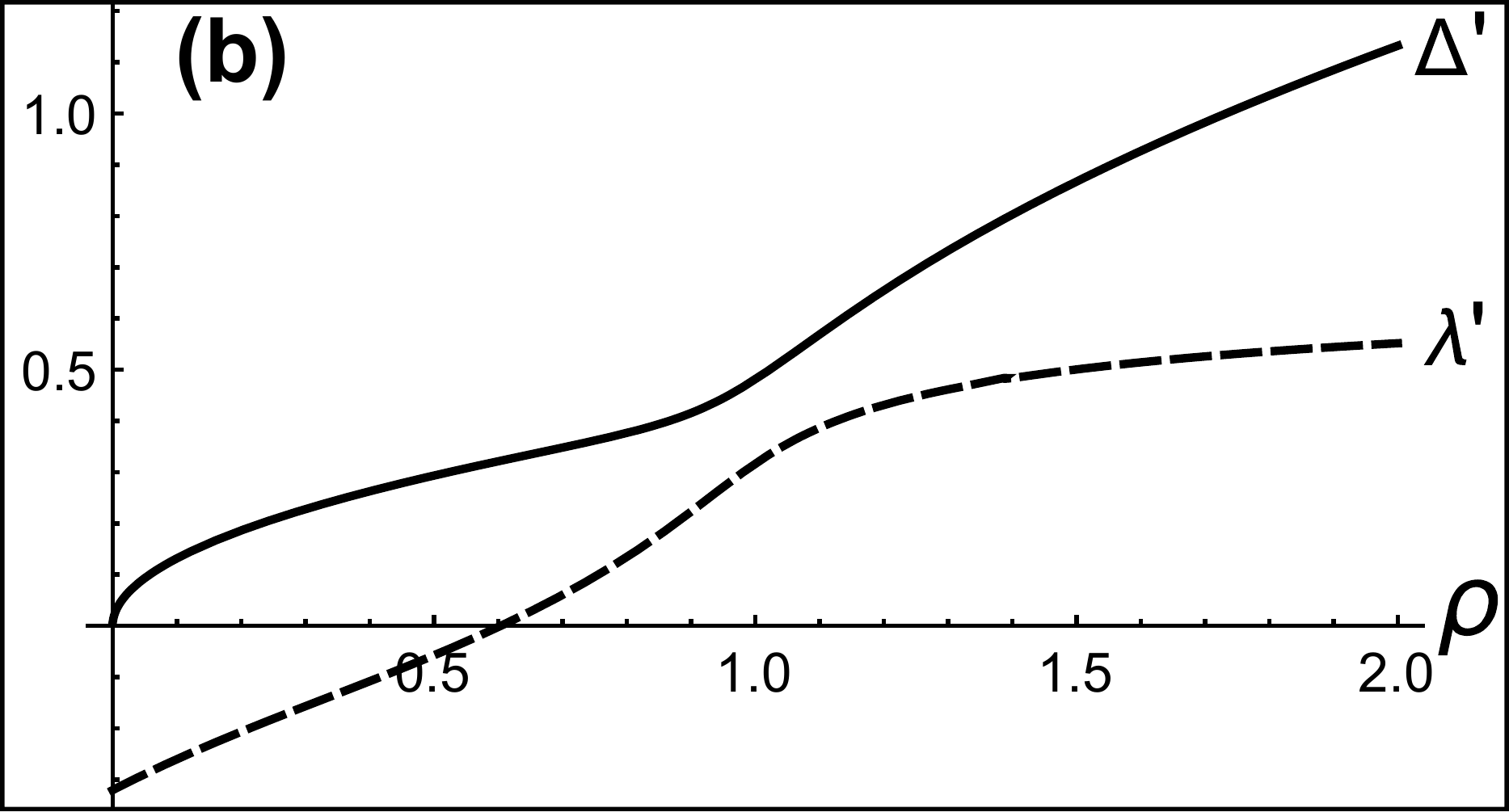}
\includegraphics[width=0.45\linewidth]{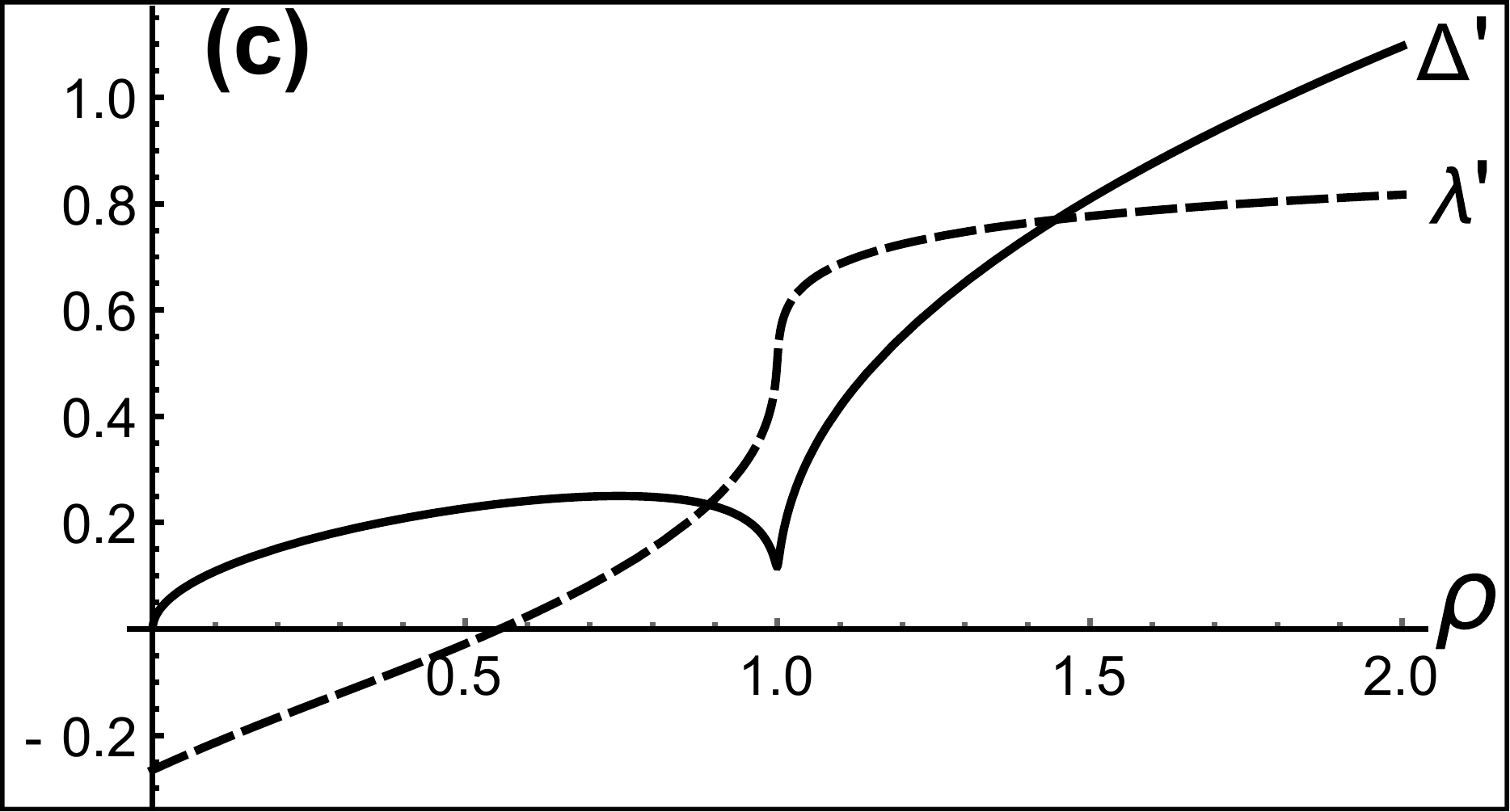}
\includegraphics[width=0.45\linewidth]{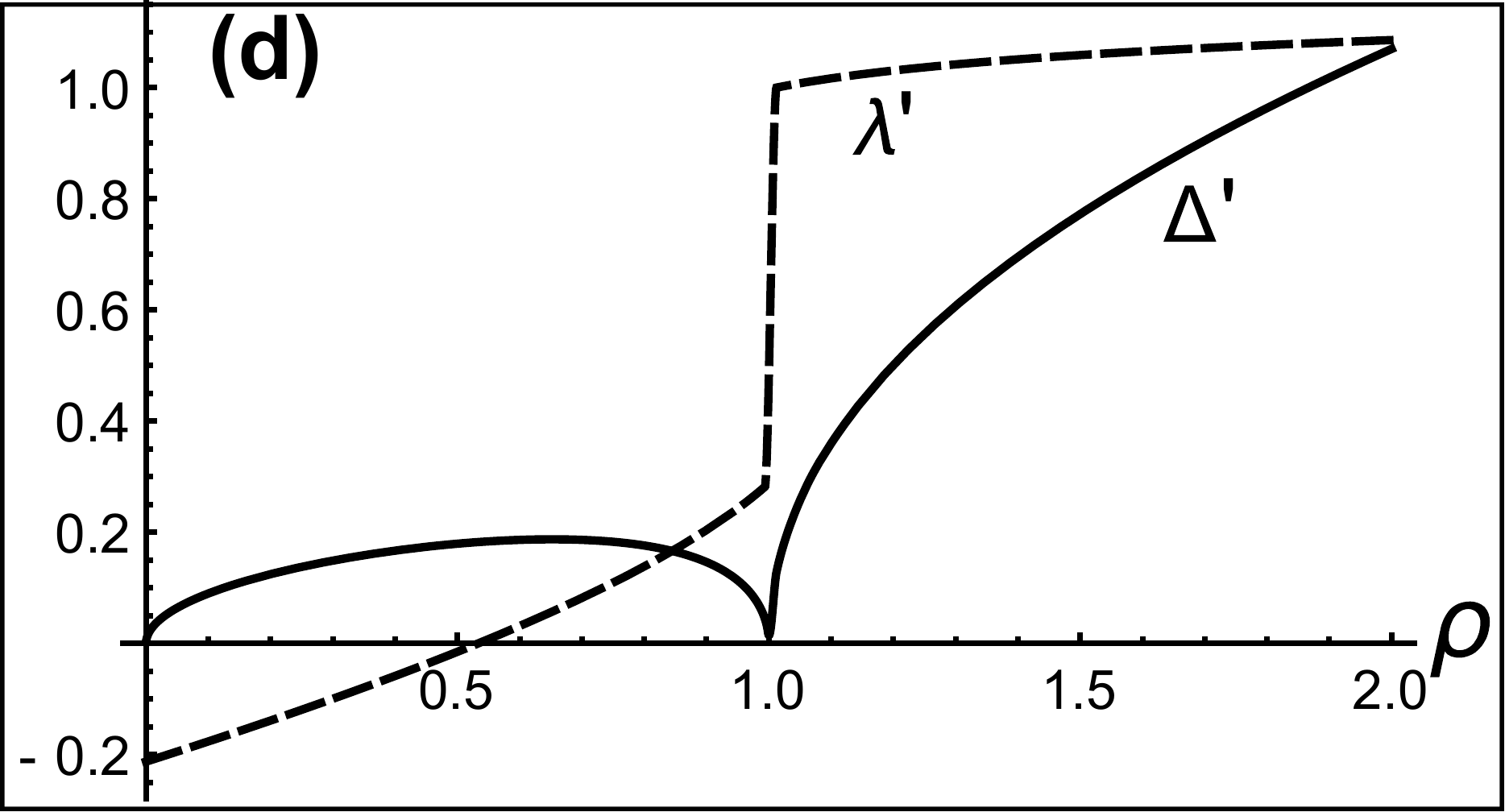}
\caption{
\label{gapchem}
The solutions of the equations for the gap $\Delta'$ (order parameter) (\ref{gap}) and chemical potential $\lambda'$ (\ref{gapchemeqs}) as functions of the pseudo-particle density $\rho$ for different values of detuning $\omega'$: $-1$ (a), 1.5 (b), 2 (c), and 2.5 (d); $\Omega'=0.3$.}
\end{figure}

Fig. \ref{gapchem} shows $\Delta'$ and $\lambda'$ as functions of pseudo-particle density $\rho = M/L$ at fixed $\Omega'=0.3$ and four different values of the detuning $\omega'=-1$ (a), 1.5 (b), 2 (c), and 2.5 (d). In all these cases, $\Delta'=0$ at $\rho=0$. At negative $\omega'$, $\Delta'$ grows monotonously as a function of $\rho$. The same behavior is revealed in the case of zero detuning $\omega'$, when the interaction between spin and photon subsystems is strongest. It also survives in the domain of positive but not too large values of $\omega'$.

However, at certain value of $\omega'>0$, qualitative changes of function $\Delta'(\rho)$ are observed. There appears a plateau in  $\Delta'$ as a function of $\rho$ (minimum of the first derivative of $\Delta'$ with respect to $\rho$), which is further transformed into a local minimum of $\Delta'$ at $\rho=1$. This minimum reaches zero, $\Delta'=0$, as the detuning increases. It indicates quantum phase transition and separates two phases -- spin-like state at $\rho <1$, when most of pseudo-particles belong to the paulion subsystem, and boson-like state at $\rho >1$, when most of them belong to the boson subsystem. The same scenario is valid for other values of $\Omega'$, but $\Delta'$, in general, decreases with the increase of broadening. Note that a spin-like state at large detuning and $\rho=1/2$ is similar to the ground state of the BCS pairing Hamiltonian. Configurations with arbitrary $\rho$ for BCS pairing Hamiltonian were analyzed in Ref. \cite{We,Pogosov,Bork}.

We would like also to notice that a similar behavior for $\Delta'$ and $\lambda'$ was reported in Ref. \cite{Littlewood} for Gaussian distribution of spin energies supplemented by cut-offs. A nontrivial evolution of both $\Delta'$ and $\lambda'$ in the vicinity of the point $\rho=1$ at large detunings is also reproduced for Gaussian distribution. Thus, the results of this subsection are qualitatively very similar to the results of Ref. \cite{Littlewood} being obtained by a different approach.

\subsection{Gap in the energy spectrum}

In reality, gap $2 \Delta_{real}'$ in the energy spectrum, i.e., energy difference between the first excited state and the ground state, is not necessarily given by $2 \Delta'$. The energy of the state with one excitation is $2\sqrt{(x_l-\lambda)^2 + \Delta^2}/g\sqrt{L}$, where $x_l$ is a given solution of the equation $F_0(z)=0$, $F_0$ being defined by (\ref{F0ansatz}). It is seen from this equation that $x_l$ is a quasi-continuous variable, which can take any value confined between $E_1$ and $E_2$, because $1/(z - \epsilon_n)$ changes from $- \infty$ to $+\infty$ when crossing the pole at $z= \epsilon_n$ along the real axis. The precise form of the quasi-continuous solution is found in Appendix E and it is used below to evaluate finite-size corrections. There is, however, an additional solution separated from quasi-continuous set of roots $x_{max} > E_2$, which is also found in Appendix E. Let us stress that this additional root, which does not exist in the case of Richardson model, is essential for a correct description of our system even in leading order in $1/L$.

\begin{figure}[h]
\includegraphics[width=0.45\linewidth]{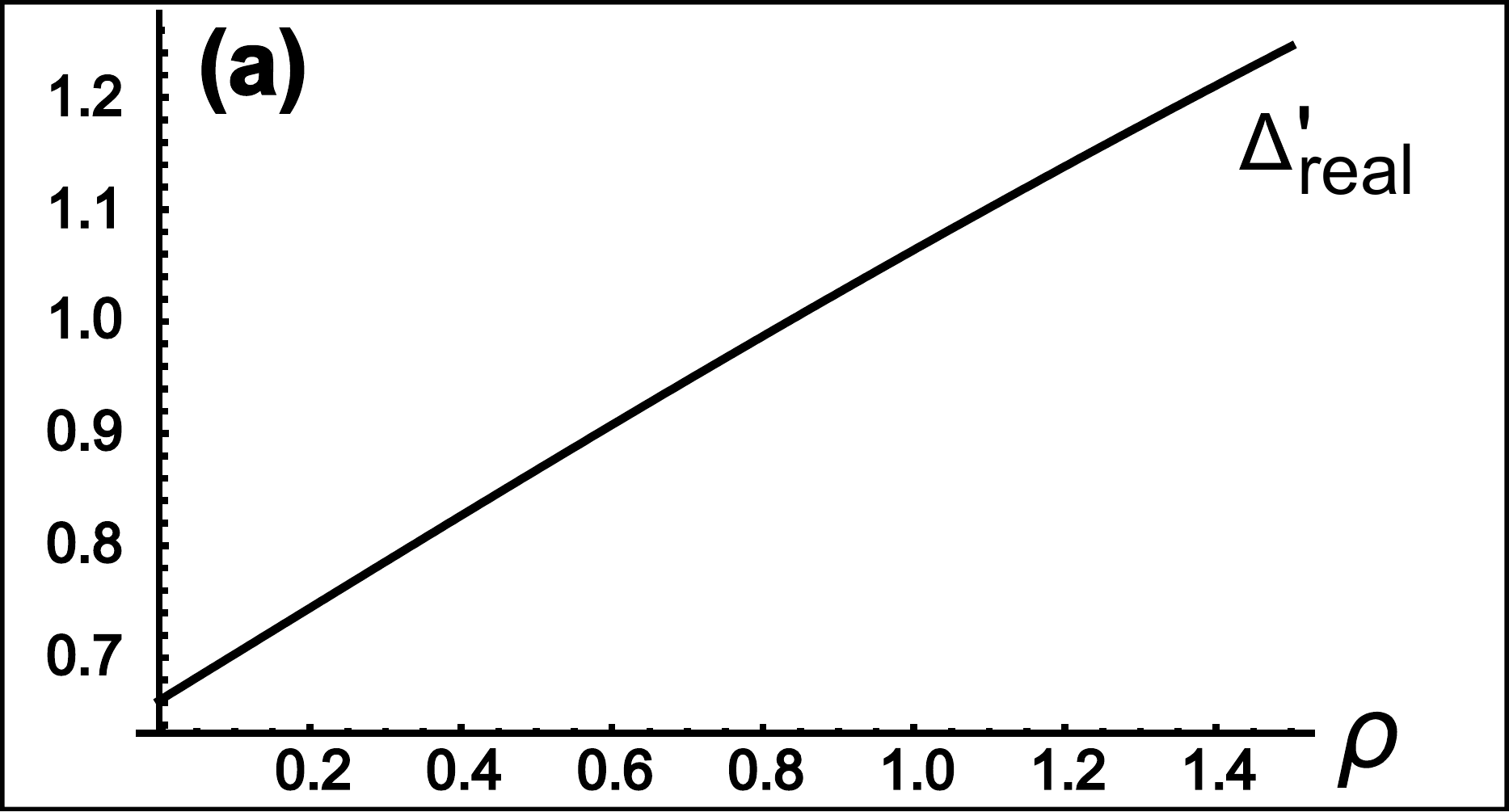}
\includegraphics[width=0.45\linewidth]{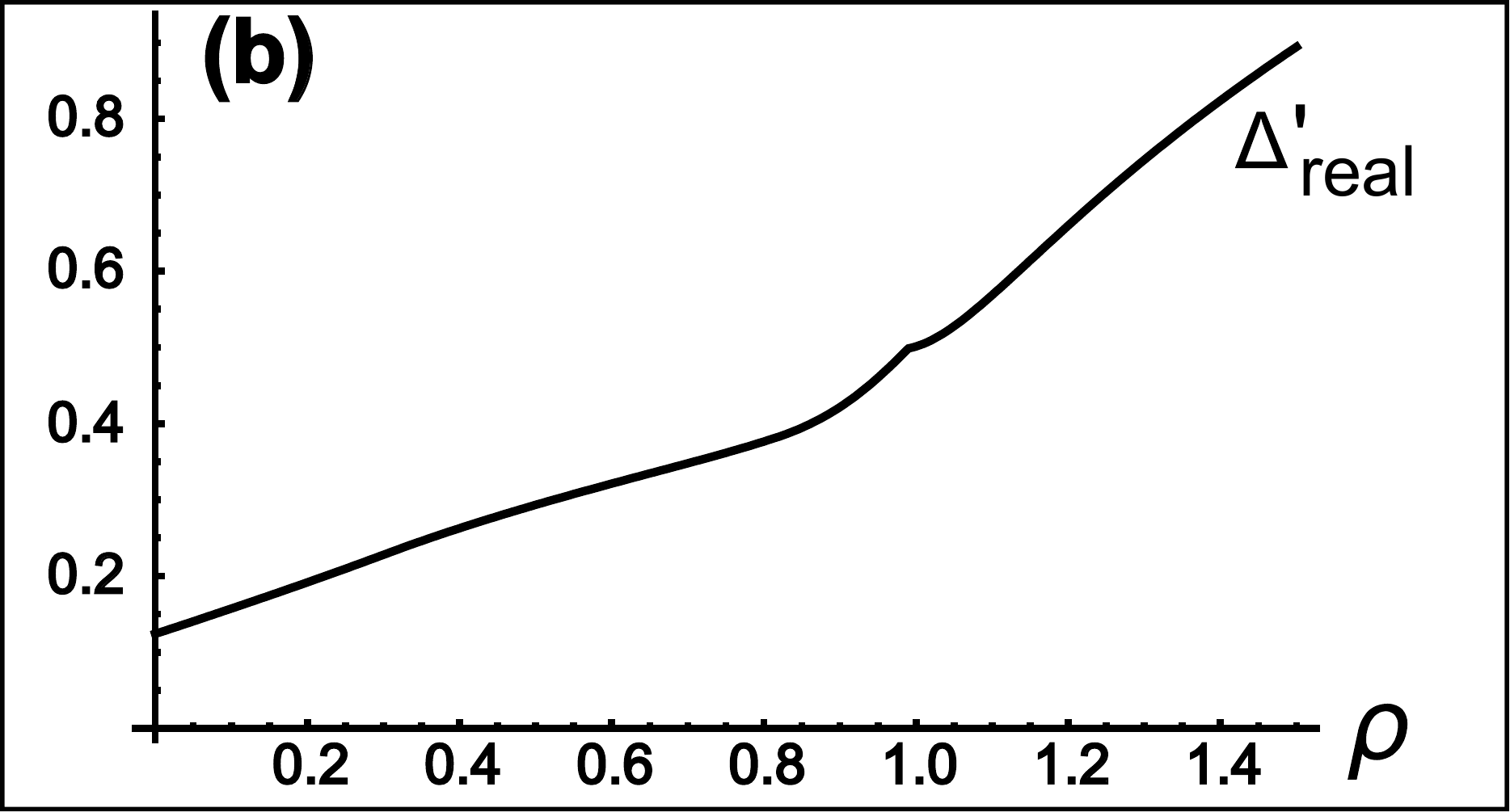}
\includegraphics[width=0.45\linewidth]{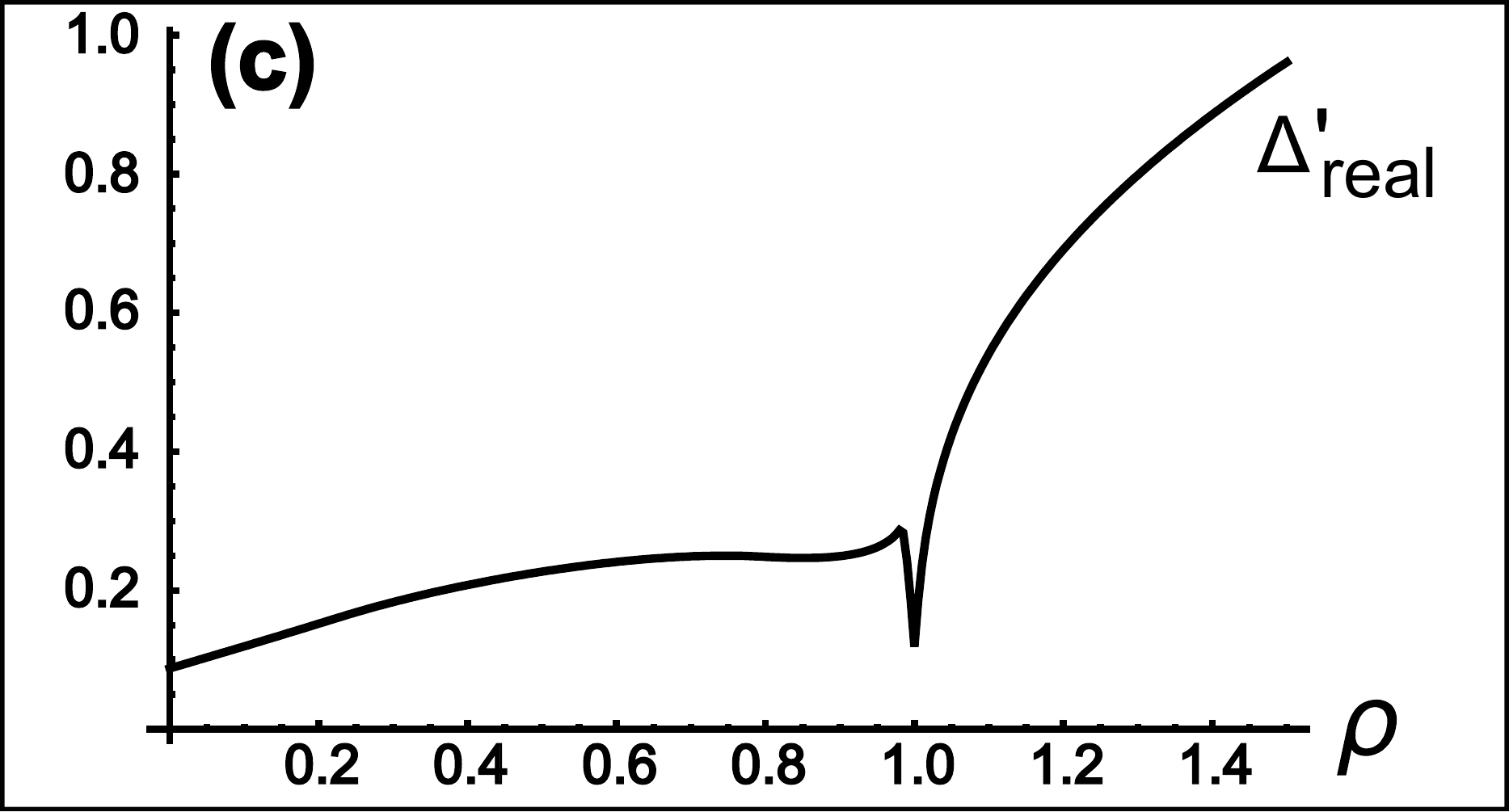}
\includegraphics[width=0.45\linewidth]{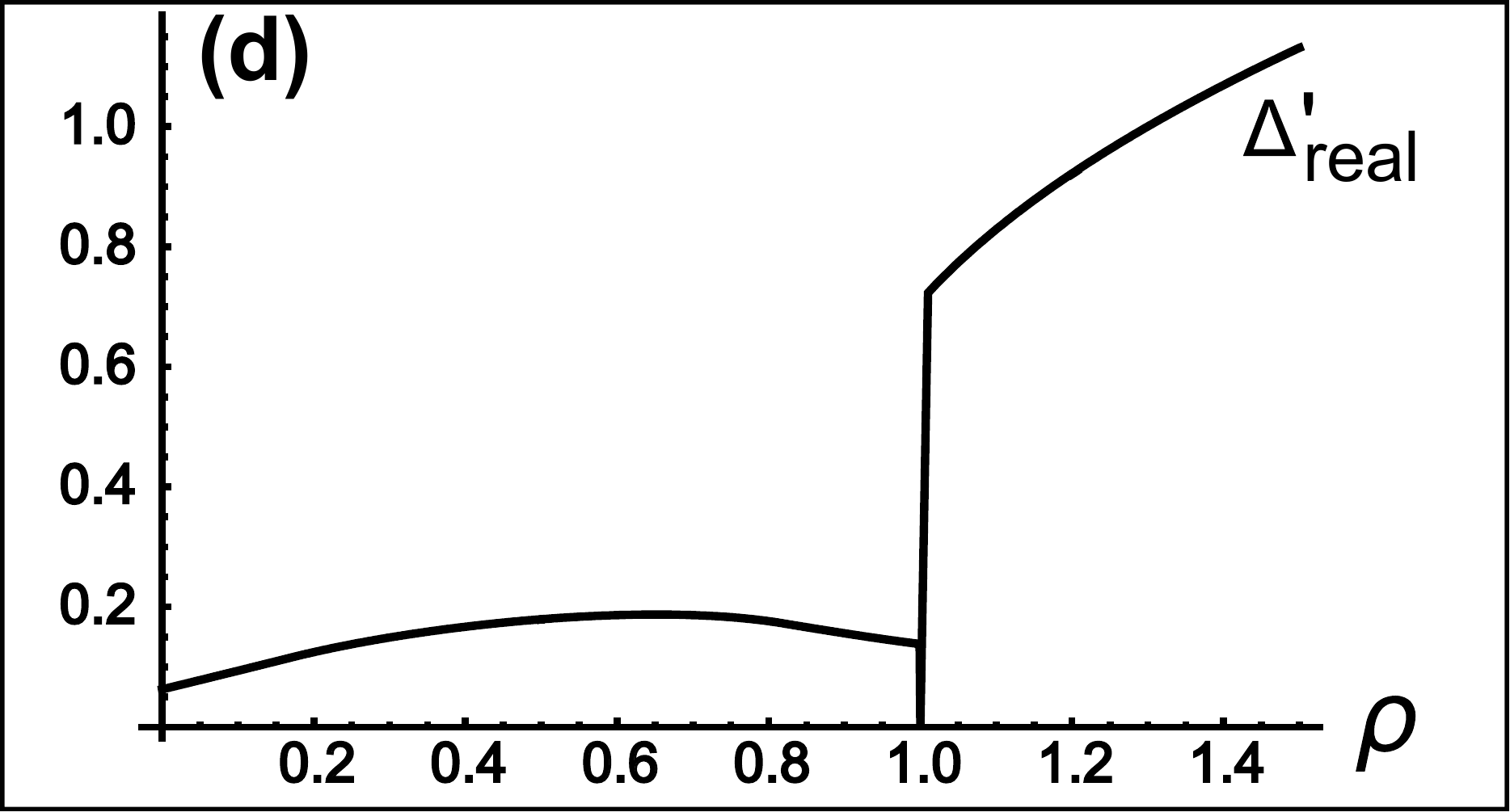}
\caption{
\label{gapmin}
Gap in energy spectrum $\Delta_{real}'$ as a function of the pseudo-particle density $\rho$ for different values of detuning $\omega'$: $-1$ (a), 1.5 (b), 2 (c), and 2.5 (d);  $\Omega'=0.3$.}
\end{figure}

There exist several scenarios for the energy gap $2 \Delta_{real}'$. The most trivial one corresponds to the configuration with $\lambda' \in [-\Omega'/2, \Omega'/2]$ (chemical potential resides between the two cut-offs). In this case, it is possible to find $x_l$ coincident with $\lambda$ up to the term $< d$, which yields an absolute minimum of $2\sqrt{(x_l-\lambda)^2 + \Delta^2}/g\sqrt{L}$ equal to $2 \Delta'$ up to a similar correction. Thus, $ \Delta_{real}'=\Delta'$ in this case.

The situation is different, provided chemical potential is below the lower cut-off, $\lambda' < -\Omega'/2$. In this case, the minimum excitation energy is attained at $x_l$ placed at this  cut-off (conditional minimum), so that $\Delta_{real}' = \sqrt{(-\Omega'/2-\lambda')^2 + \Delta'^2}$.

There is also a possibility that chemical potential is above the upper cut-off, $\lambda' > -\Omega'/2$. In this case, we have to find excitation energies corresponding to $x_l$ placed at the upper cut-off and at $x_{max}$ and then to compare them. The lowest of the two quantities provides $2 \Delta_{real}'$. Thus, we have to choose a minimum from  $\sqrt{(\Omega'/2-\lambda')^2 + \Delta'^2}$ and $\sqrt{(x_{max}'-\lambda')^2 + \Delta'^2}$.

\begin{figure}[h]
\includegraphics[width=0.45\linewidth]{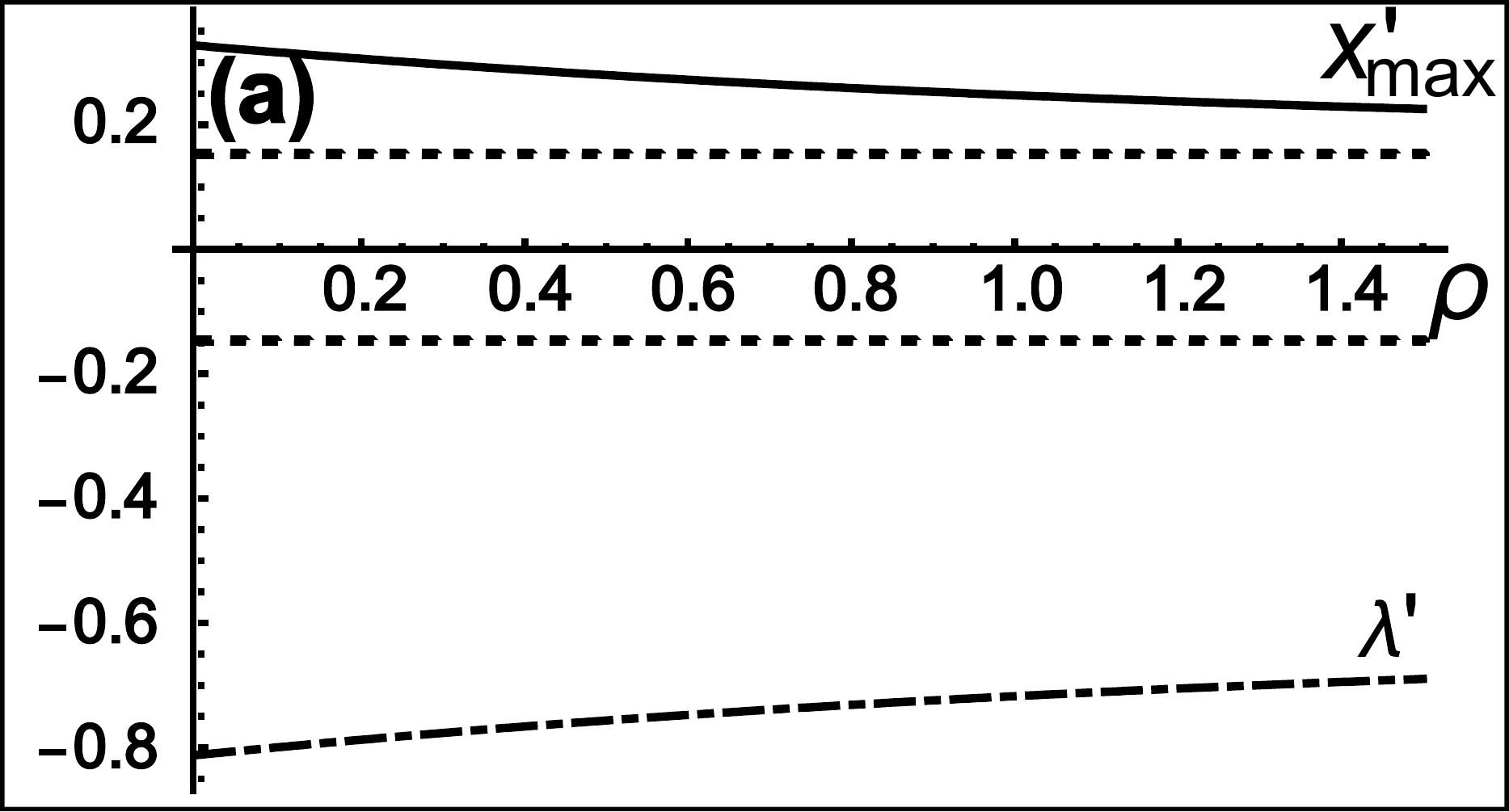}
\includegraphics[width=0.45\linewidth]{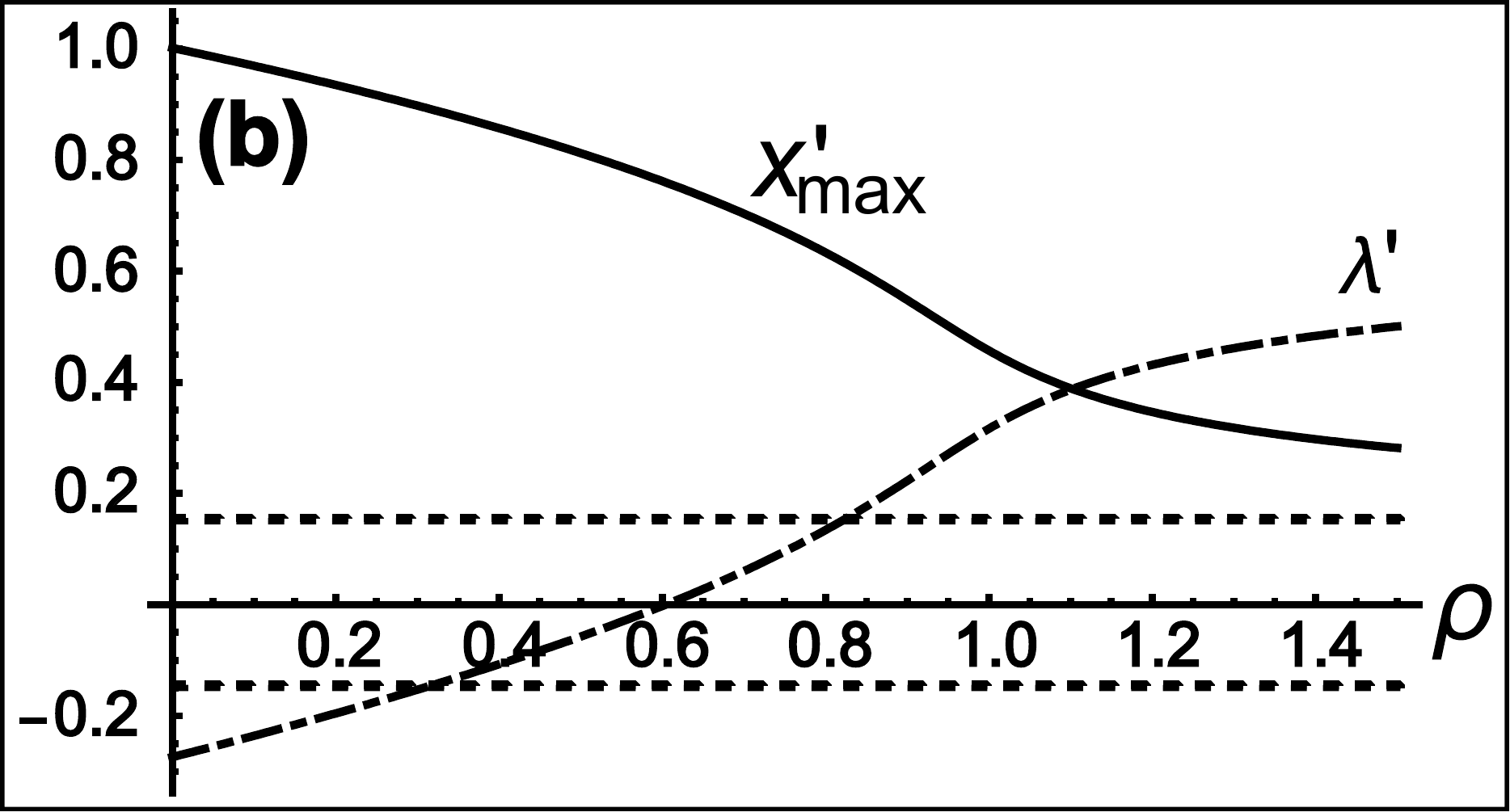}
\includegraphics[width=0.45\linewidth]{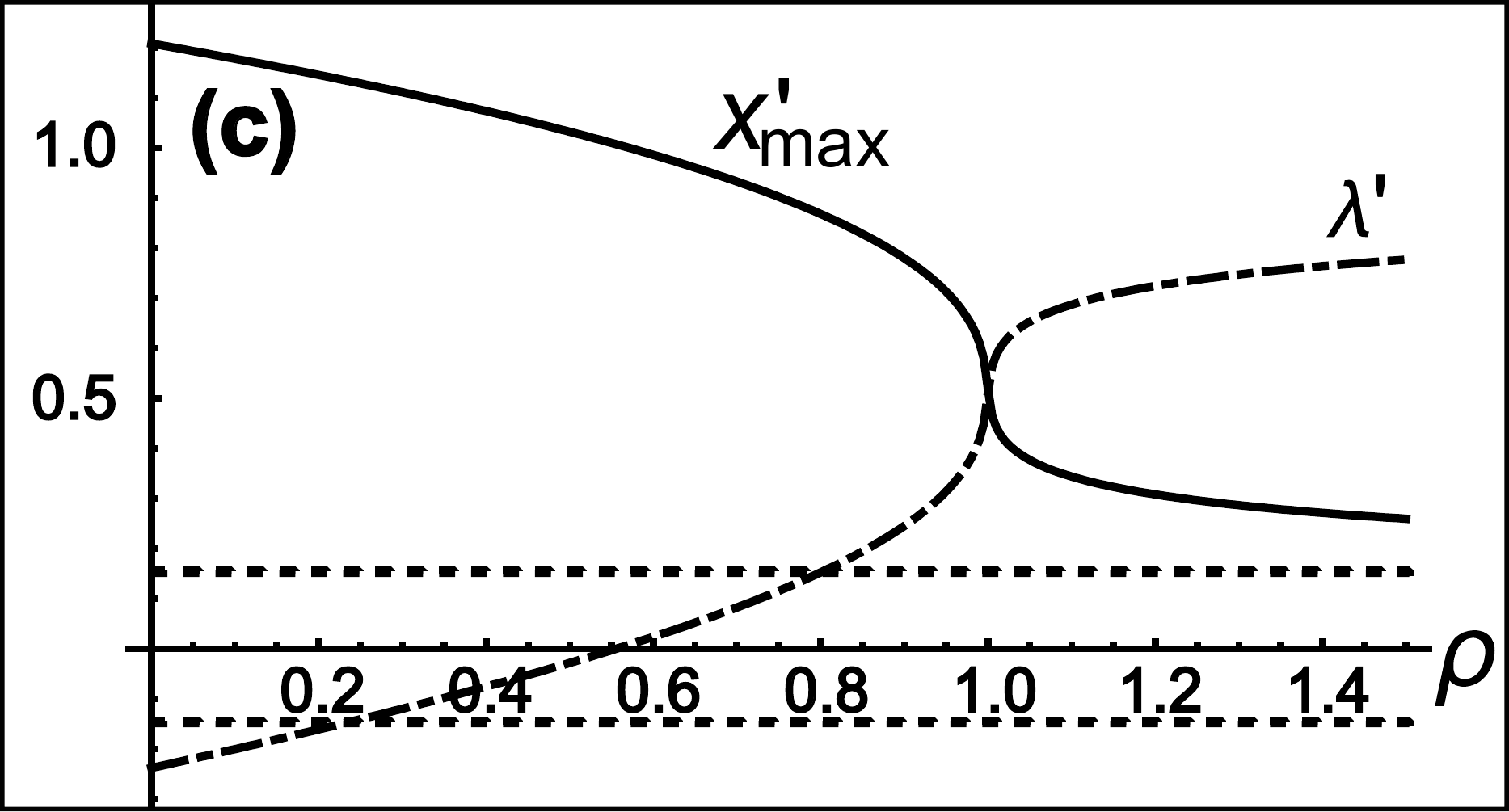}
\includegraphics[width=0.45\linewidth]{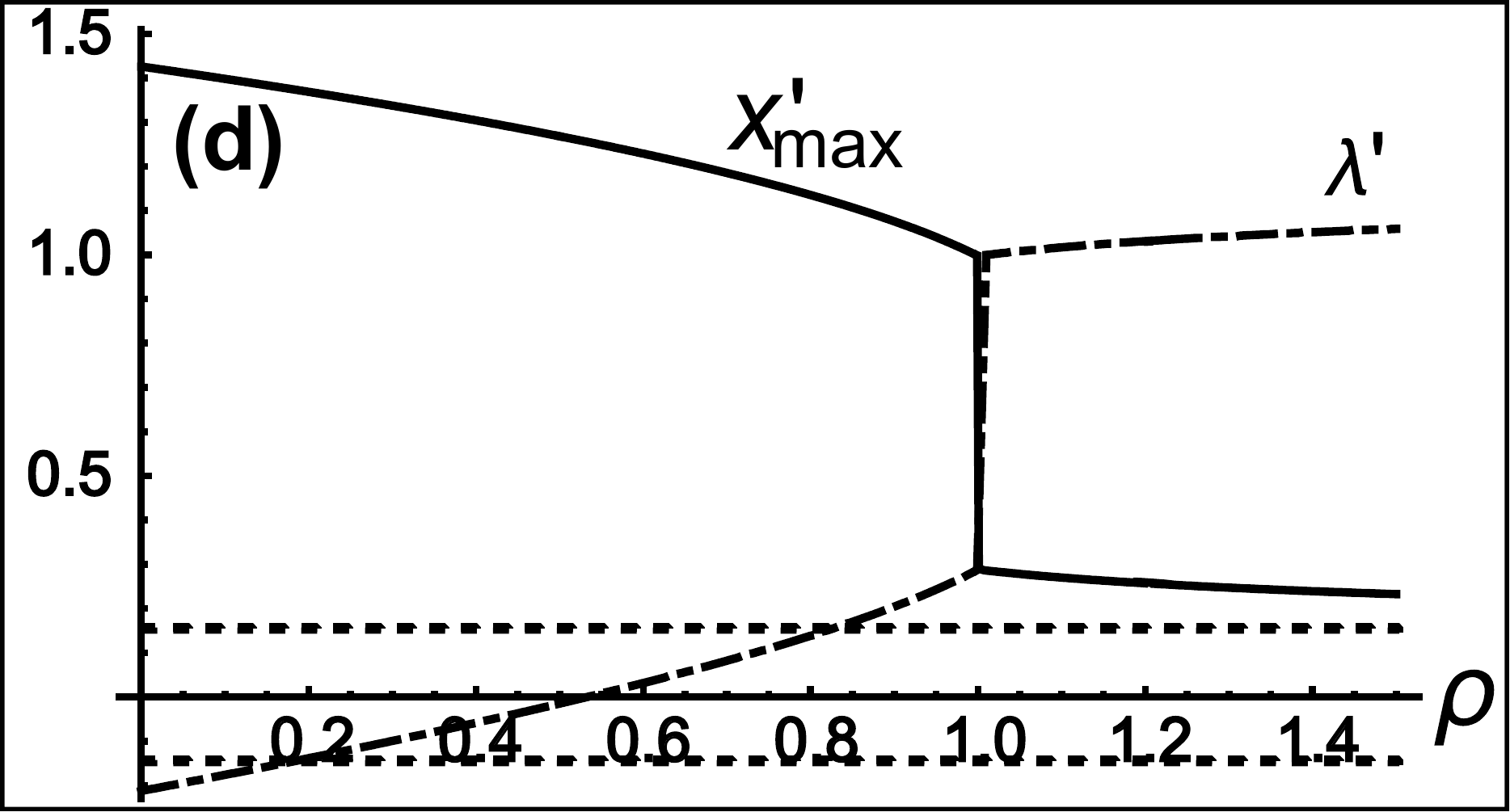}
\caption{
\label{xmax}
The dependencies of $x_{max}'$ (solid line) and $\lambda'$ (dash-dot line) on the pseudo-particle density $\rho$ for different values of detuning $\omega'$: $-1$ (a), 1.5 (b), 2 (c), and 2.5 (d); $\Omega'=0.3$. Also shown are the endpoints of the interval, which confines spin energies (dot line).}
\end{figure}

\begin{figure}[h]
\includegraphics[width=0.45\linewidth]{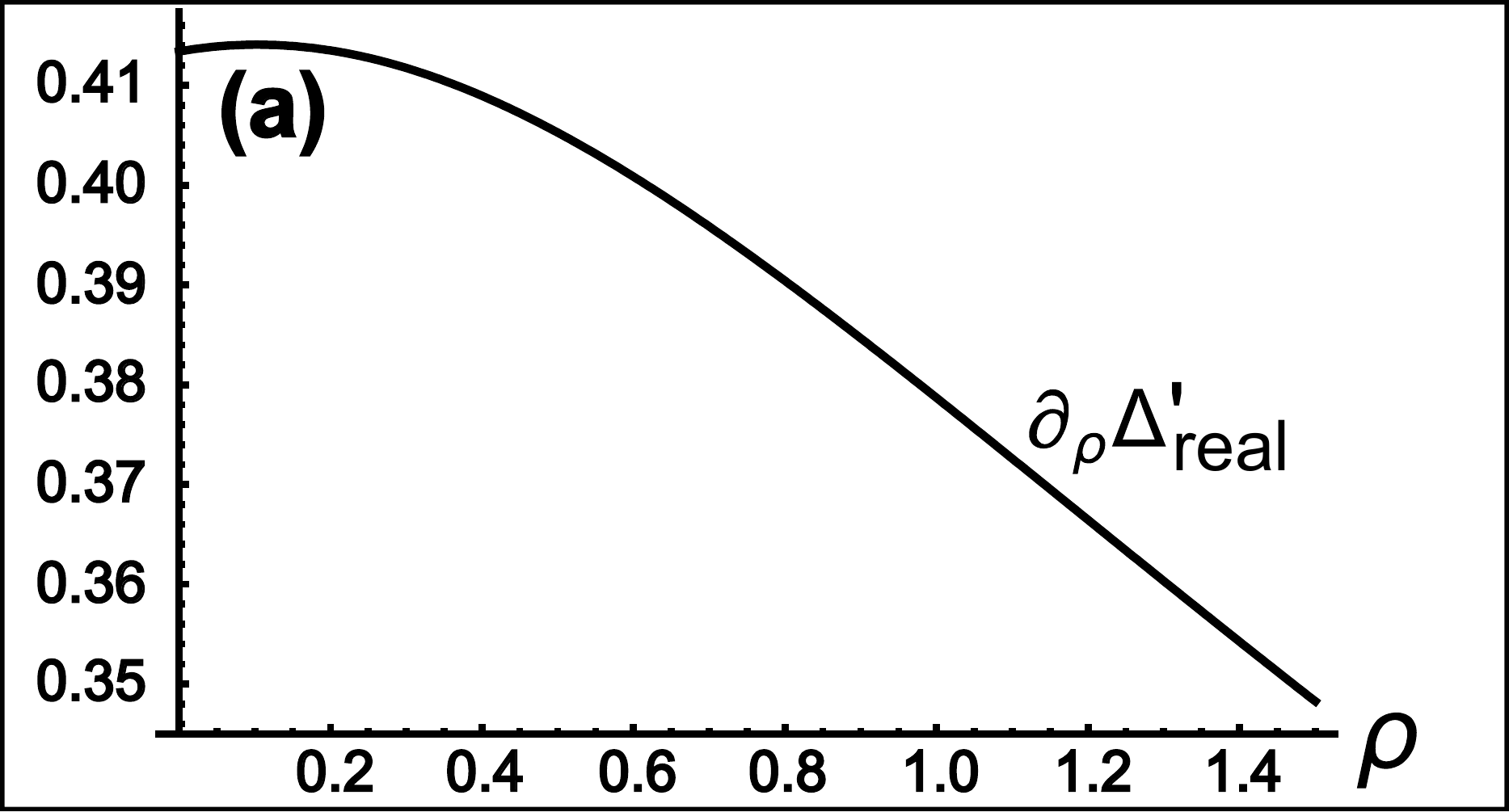}
\includegraphics[width=0.45\linewidth]{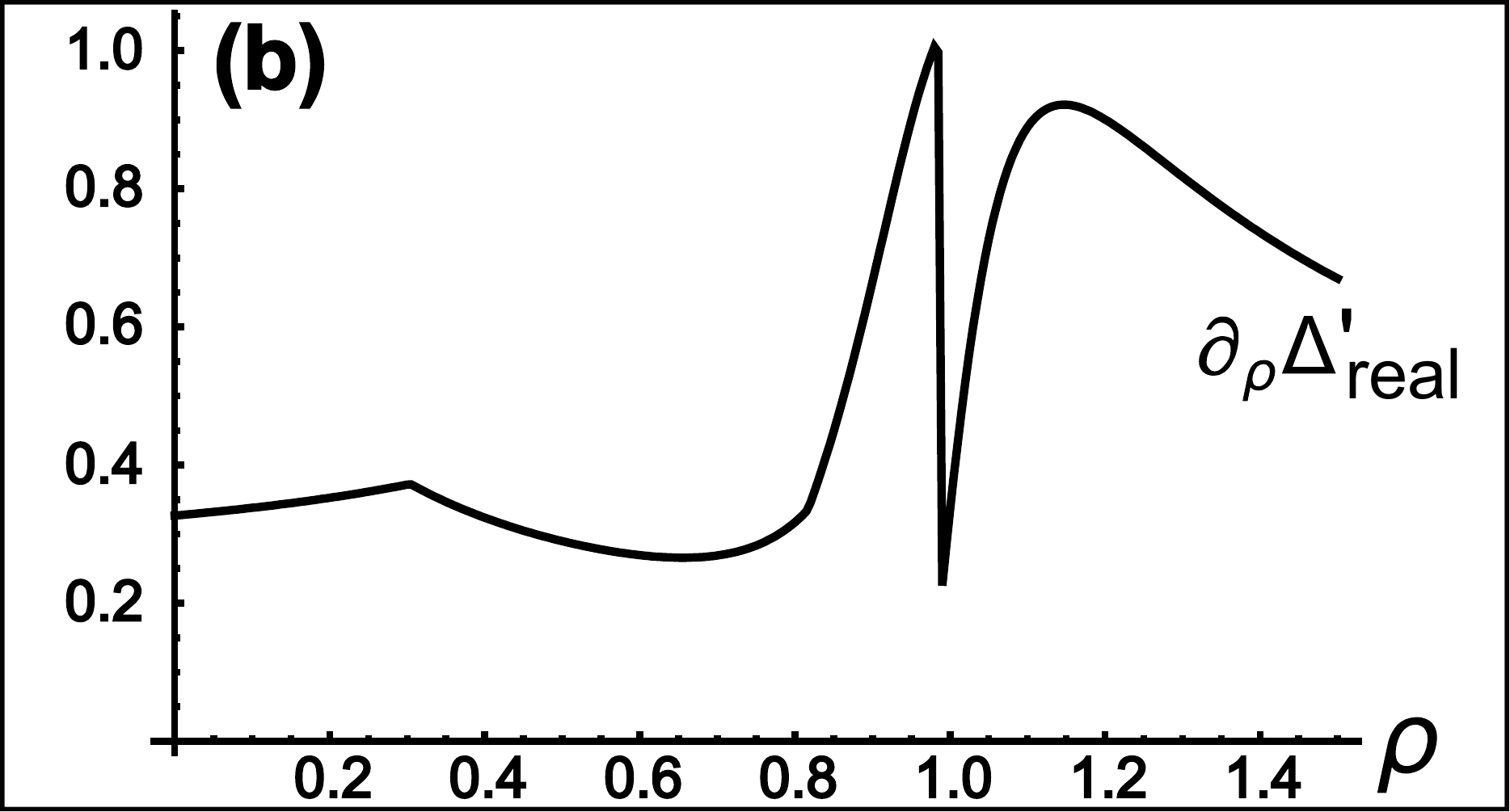}
\includegraphics[width=0.45\linewidth]{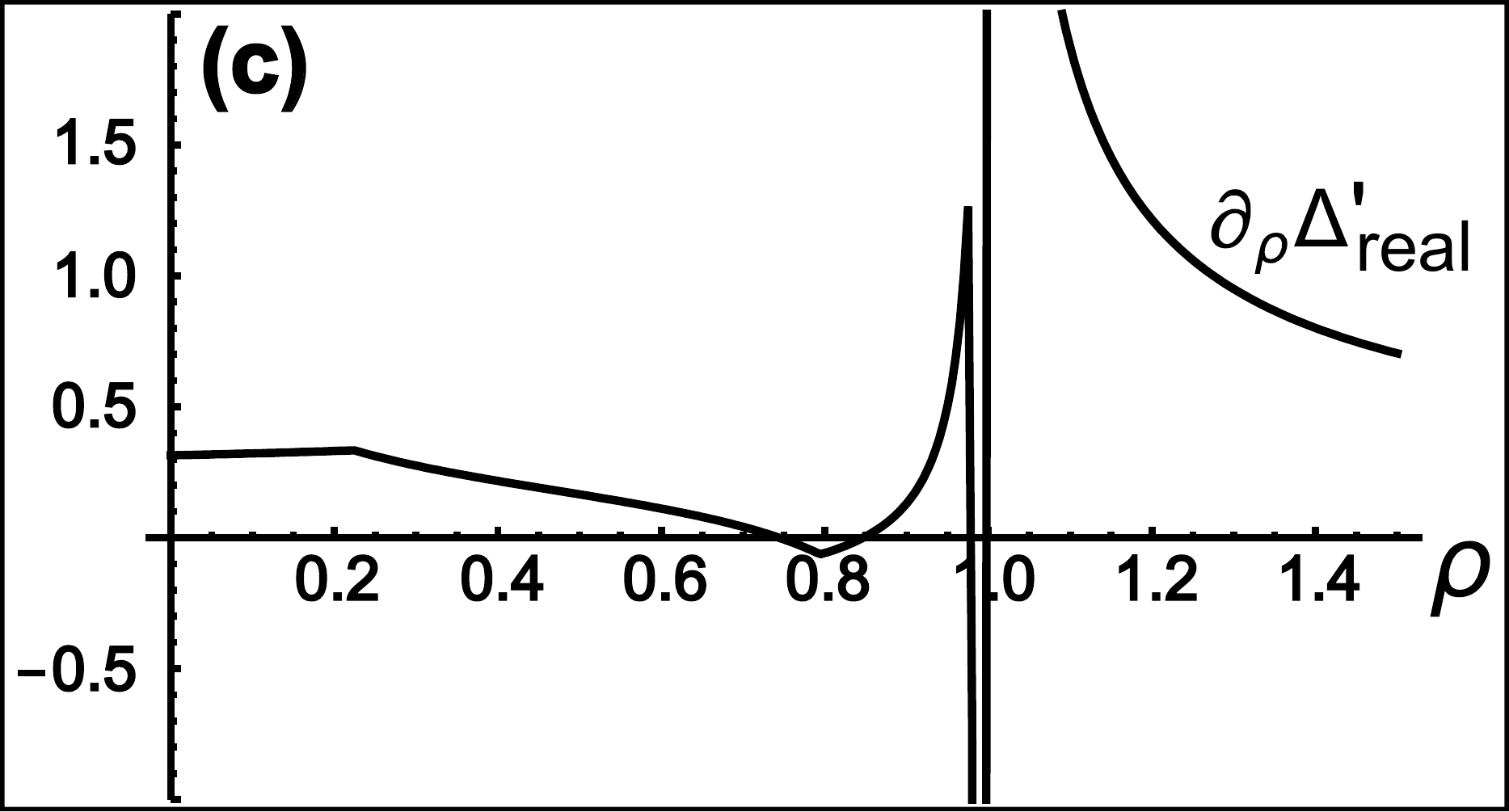}
\includegraphics[width=0.45\linewidth]{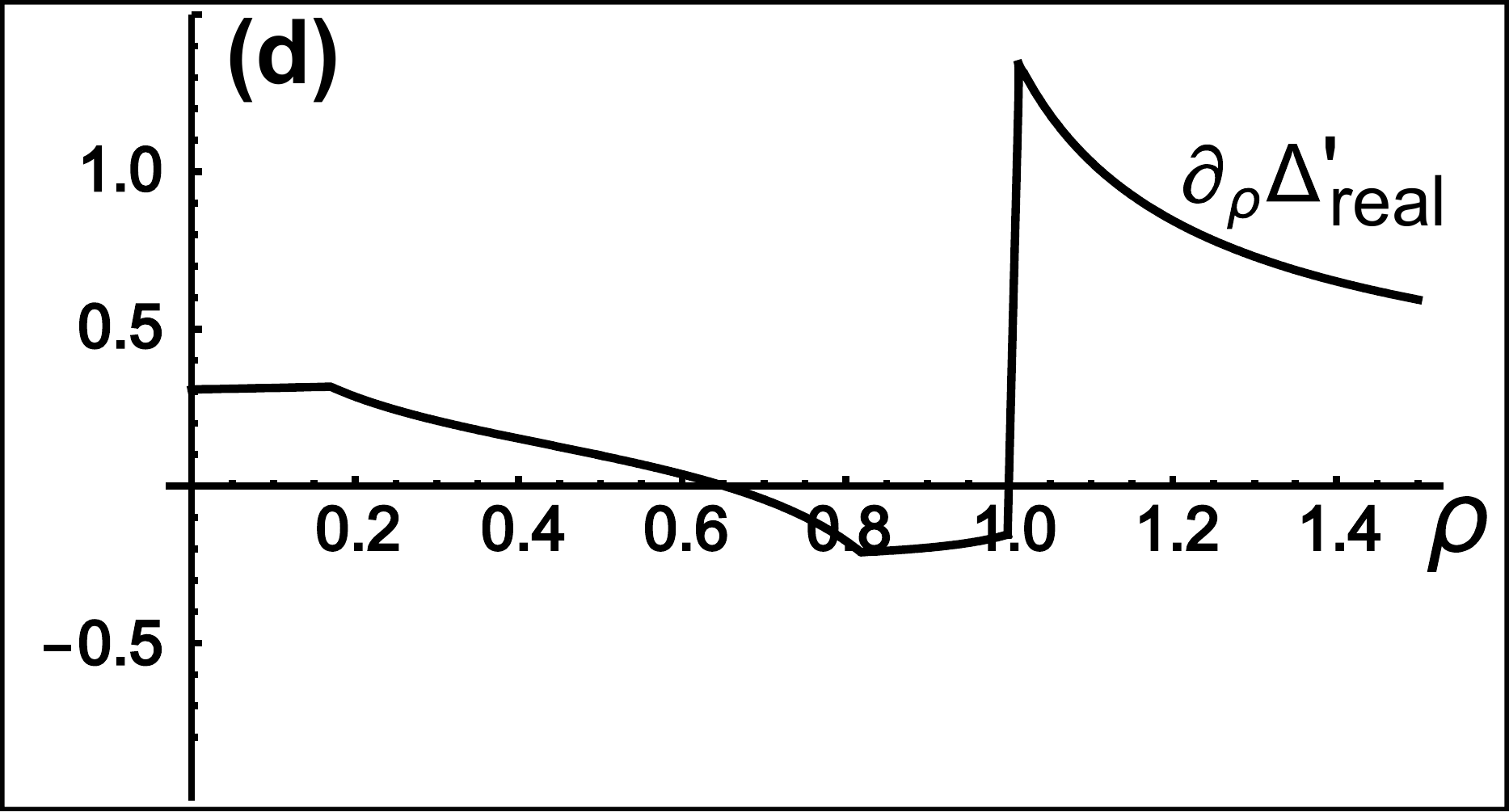}
\caption{
\label{deriv}
First derivative of $\Delta_{real}'$ with respect to $\rho$ as a function of $\rho$ for different values of detuning $\omega'$: $-1$ (a), 1.5 (b), 2 (c), and 2.5 (d); $\Omega'=0.3$.}
\end{figure}

The result of our numerical calculations for $\Delta_{real}'$ as a function of $\rho$ is presented in Fig. \ref{gapmin} for $\Omega'=0.3$ and four different values of the detuning $\omega'=-1$ (a), 1.5 (b), 2 (c), and 2.5 (d). The behavior of $\lambda'$ and $x_{max}'$ is shown in Fig. \ref{xmax}. In order to visualize peculiarities of function $\Delta_{real}'(\rho)$, we also plot in Fig. \ref{deriv} the first derivative $d \Delta_{real}' / d \rho $, which is denoted as $\partial \Delta_{real}'$.

We find that $\lambda'$ is always below the lower cut-off at $\omega' < 0$ (Fig. \ref{xmax} (a)). In this case, $\Delta_{real}'$ is a monotonously increasing function of $\rho$, as Fig. \ref{gapmin} (a) illustrates. Gap in the energy spectrum is determined by a joint contribution of $\lambda'$ and $\Delta'$ and, therefore, is nonzero at $\rho=0$ in contrast to $\Delta'$. No discontinuity appears in $\partial \Delta_{real}'$, see Fig. \ref{deriv} (a). In this case, most of pseudo-particles are in a boson state, which nevertheless interact with each other via spin subsystem.

Chemical potential $\lambda'$ starts to enter the interval $[-\Omega'/2, \Omega'/2]$, as $\omega'$ increases. This is illustrated by Fig. \ref{xmax} (b). The entrance is accompanied by the change in the behavior of function $\Delta_{real}'$ at corresponding value of $\rho$. The change is not discernible in Fig. \ref{gapmin} (b), where it occurs at $\rho \approx 0.35$, but is clearly seen in Fig. \ref{deriv} (b), since it signals as a discontinuity in the second derivative of $\Delta_{real}'$. At larger density $\rho \approx 0.83$, $\lambda'$ starts to fall above the interval $[-\Omega'/2, \Omega'/2]$. It turns out that $\Delta_{real}'$ at $\rho \gtrsim 0.83$ corresponds to $x_l$ placed at upper cut-off and not at $x_{max}$. A discontinuity in the second derivative of $\Delta_{real}'$ at $\rho \approx 0.83$ is again poorly discernible in Fig. \ref{gapmin} (b), but it is visible in Fig. \ref{deriv} (b). At $\rho=1$, there appears a discontinuity of the first derivative of $\Delta_{real}'$ in Fig. \ref{deriv} (b). This is due to the fact that the energy gap at this point starts to be associated with $x_{max}'$ and not with the upper cutoff. Slightly above $\rho=1$, $\lambda'$ and $x_{max}'$ approach each other until they collide, but they start to move away from each other at larger values of $\rho$. This process is shown in Fig. \ref{xmax} (b). It is accompanied by the nonmonotonous behavior of the first derivative of $\Delta_{real}'$, as it is shown in Fig. \ref{deriv} (b). All these peculiarities can hardly be extracted from Fig. \ref{gapmin} (b).

Further increase of $\omega'$ is accompanied by the appearance of the local minimum of $\Delta'(\rho)$ at $\rho=1$ (Fig. \ref{gapchem} (c)). In this case, the behavior of $x_{max}'$ and $\lambda'$, which is shown in Fig. \ref{xmax} (c), is qualitatively the same as in the previously described situation. However, local minimum in $\Delta'(\rho)$ now gives rise to the local minimum of $\Delta'(\rho)$ at $\rho=1$, clearly visible in Fig. \ref{gapmin} (c).

Let us now consider larger values of $\omega'$, which support a full suppression of $\Delta'$ to zero at $\rho=1$. In this case, at $\rho=1$, both $x_{max}'$ and $\lambda'$ experience a discontinuity by jumping towards each other, as illustrated in Fig. \ref{xmax} (d). Because order parameter $\Delta'$ vanishes in this highly singular point, spectral gap $\Delta_{real}'$ vanishes as well. Due to the jump in both $x_{max}'$ and $\lambda'$, $\Delta_{real}'$ experiences a discontinuity, clearly visible in Fig. \ref{gapmin} (d). A step-like behavior of the gap can be understood by considering a limit of noninteracting system. In this regime, at large detuning and at $\rho=1$ (complete filling), single excitation with lowest energy is obtained by the de-excitation of the highest-energy spin with the subsequent creation of a boson. Thus, in this case, gap in the energy spectrum should jump from 0 to $\omega'-\Omega'$ when crossing the point $\rho=1$.

Note that, as $\Omega'$ grows, the chemical potential starts to enter the interval $[-\Omega'/2, \Omega'/2]$ already at $\rho=0$. This results in vanishing of $\Delta_{real}'$ at $\rho=0$.

To the best of our knowledge, such an analysis of the 'fine structure' of the gap in energy spectrum for the inhomogeneous Dicke model has not been performed yet. For instance, Ref. \cite{Littlewood} is focused on spectral density rather than on such issues.

\subsection{Ground state energy in leading order}

Let us now analyze the ground state energy in leading order in $1/L$. This quantity can be found from Eq. (\ref{energygr0}) by replacing sums by integrals and using relations (\ref{gap}), (\ref{gapchemeqs}). After some algebra, we obtain in the dimensional units
\begin{eqnarray}
E_{gr0}=ME_1+M_{spin}^2d+\frac{2\Delta^2}{g^2}M_{spin}d-2d\frac{\sigma}{1-\sigma}\left(M_{spin}(L-M_{spin})+\frac{\Delta^2}{g^2}(L-2M_{spin})\right),
\label{Egrodimens}
\end{eqnarray}
where
\begin{eqnarray}
M_{spin}=M-\frac{\Delta^2}{g^2}
\label{Mspin}
\end{eqnarray}
is a mean number of spin pseudo-particles, while
\begin{eqnarray}
\sigma=\exp(-2\Omega'(\omega'-\mu'))
\label{sigma}
\end{eqnarray}
is a nonanalytical function of the effective interaction constant. The emergence of such functions is standard for thermodynamical limit of BCS models.

The knowledge of a ground state energy is of particular importance in the view of a superradiant transition, which is a characteristic feature of Dicke model. Namely, if the spin-boson interaction is strong enough, the ground state energy as a function of pseudo-particle density $\rho$ can have a minimum not at $\rho=0$, but at some finite density. Note, however, that the expression of the ground state energy given by (\ref{Egrodimens}) depends also on the lower cut-off $E_1$, which enters this quantity additively. It is reasonable, therefore, to define a related quantity $E_{super}$ as
\begin{eqnarray}
E_{super}=E_{gr0}-M \min (2E_1,\omega).
\label{Esuper}
\end{eqnarray}
 An additive contribution $M \min (2E_1,\omega)$ to the total energy increases linearly with the increase of $\rho$. However, the remaining contribution, $E_{super}$, can be a negative decreasing function of $\rho$. This can lead to the superradiant transition, provided $\min (2E_1,\omega)$ is small enough.

\begin{figure}[h]
\includegraphics[width=0.45\linewidth]{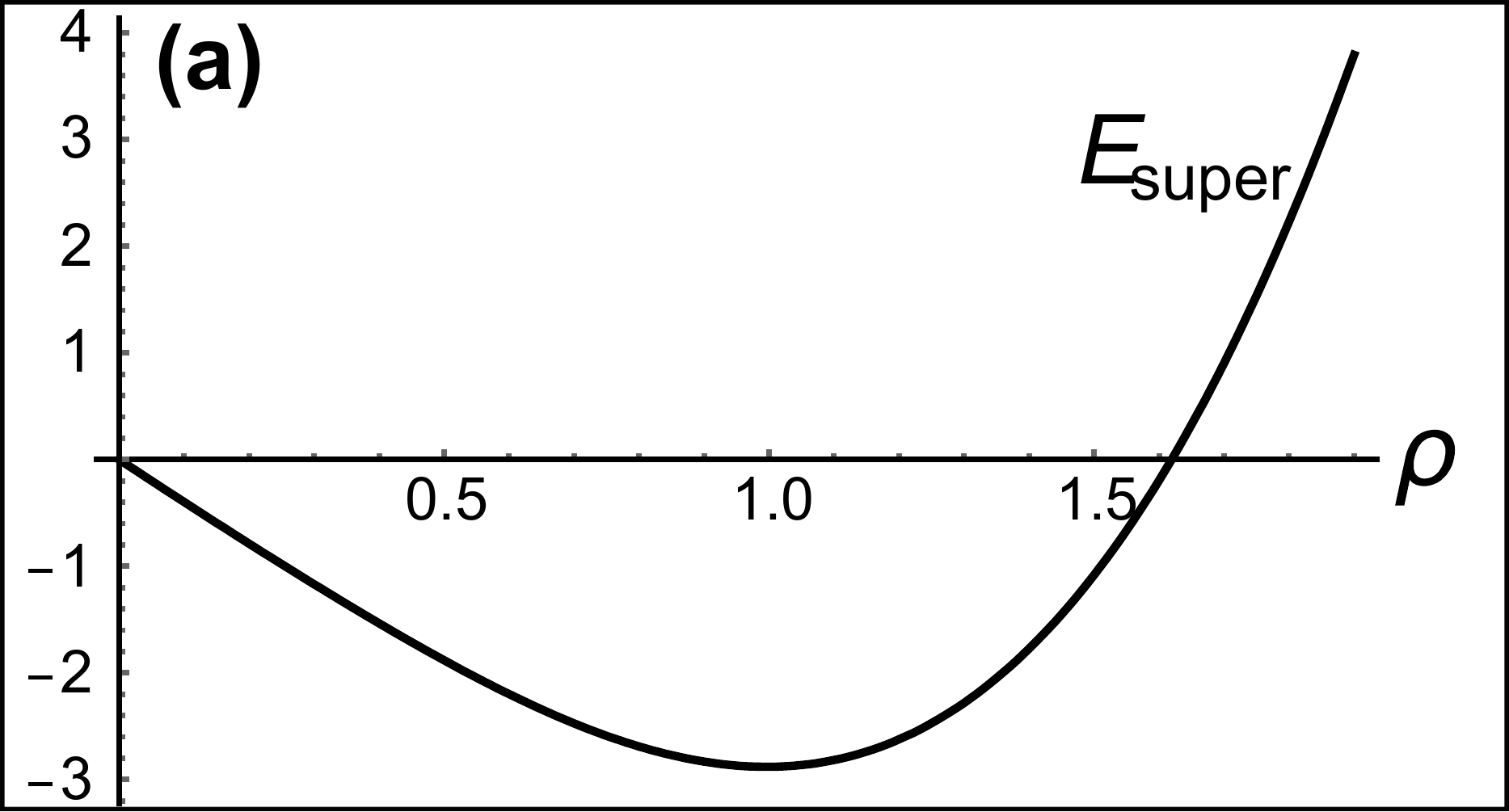}
\includegraphics[width=0.45\linewidth]{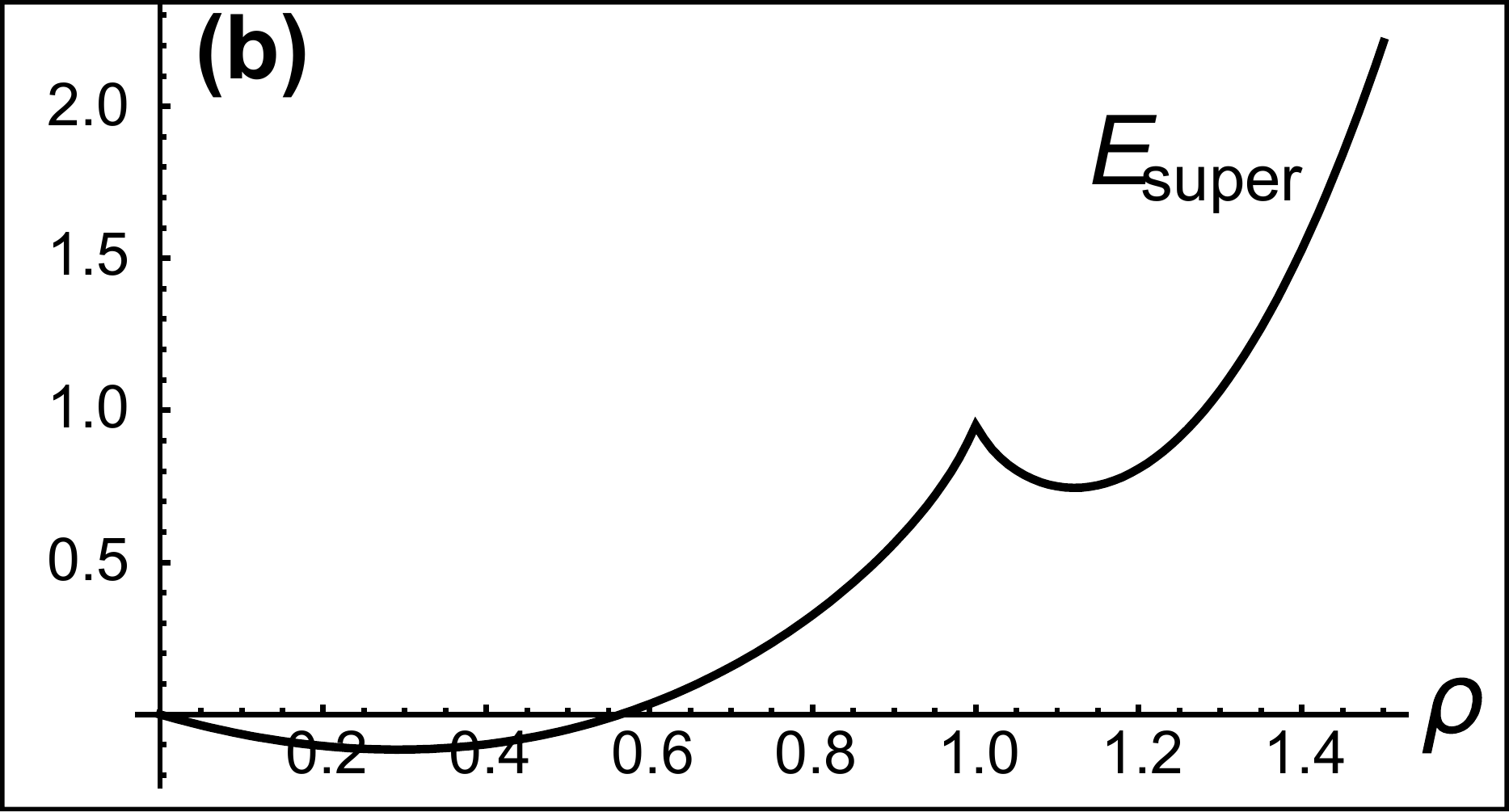}
\caption{
\label{Esuper}
The dependence of $E_{super}$ defined in Eq. (\ref{Esuper}) on pseudo-particle density $\rho$ at $\omega' = 0$ (a) and 2 (b); $\Omega'=0.3$.}
\end{figure}

In Fig. (\ref{Esuper}), we plot $E_{super}/L\Omega$ as a function of pseudo-particle density $\rho$ for $\Omega'=0.3$ and two different values of the detuning $\omega' = 0$ (a) and 2 (b). This quantity has a minimum at some nonzero $\rho$, provided the detuning is not too large. A decrease of $E_{super}/L\Omega$, as $\rho$ increases from zero, can be linear. It is able, therefore, to overcome positive contribution $M \min (2E_1,\omega)$ to the total energy and, consequently, to provoke a superradiant transition. We also find that smaller and smaller detuning $\omega'$ is needed to attain a region with negative $E_{super}/L\Omega$, as inhomogeneous broadening $\Omega'$ increases. This fact illustrates a negative role played by both the inhomogeneous broadening and detuning in the hybridization of the spin and boson subsystems.

Another important quantity, which can be extracted from the ground state energy, is an interaction energy $E_{inter}$. This is the difference between the total ground state energy $E_{gr0}$ and the energy $E_{gr0}^{(nonint)}$ of the same number $M$ of noninteracting pseudo-particles calculated to the same accuracy in $1/L$. The latter must be evaluated with special care, because several distinct situations do exist. Below we briefly describe them.

\begin{figure}[h]
\includegraphics[width=0.3\linewidth]{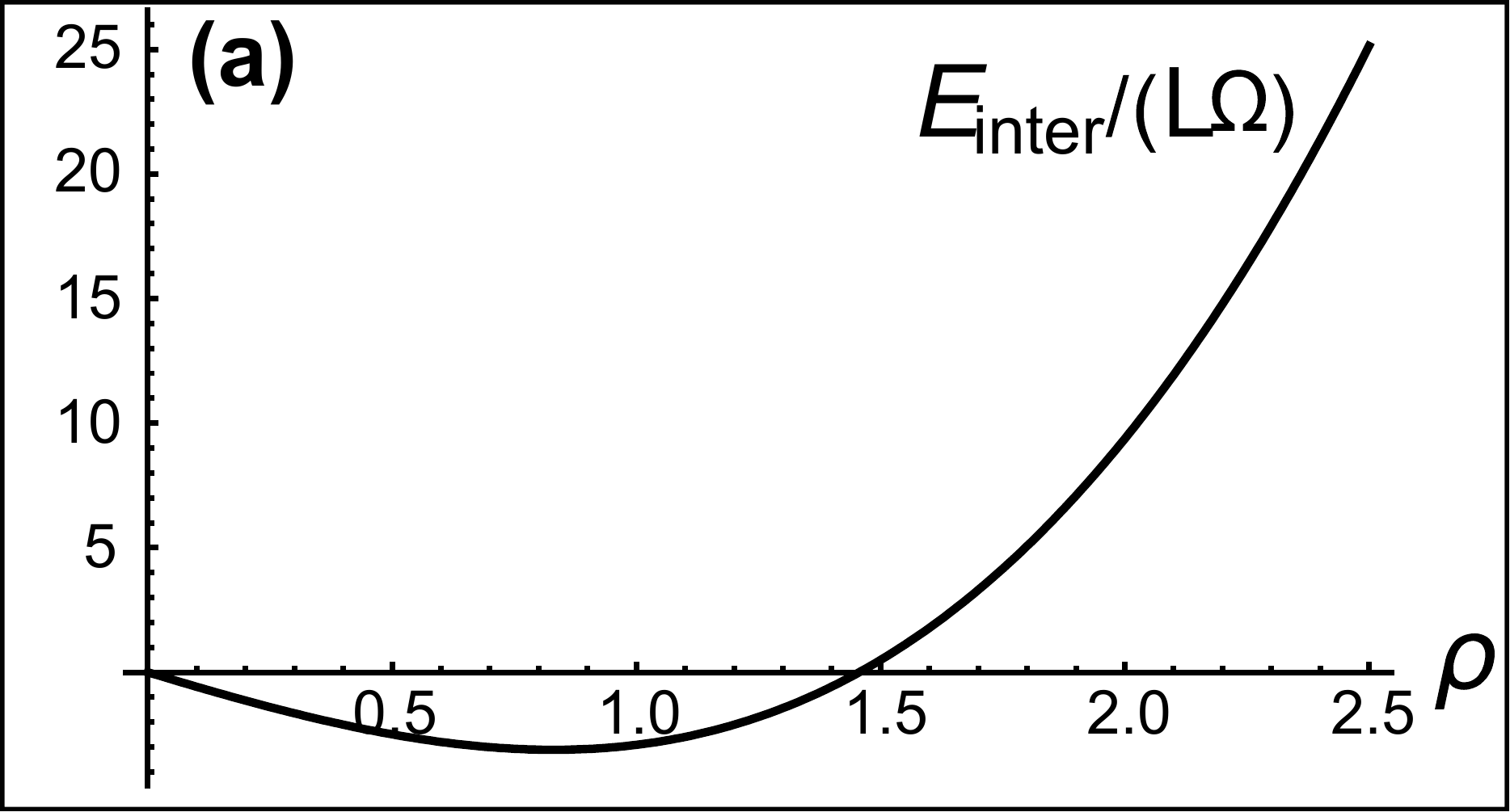}
\includegraphics[width=0.3\linewidth]{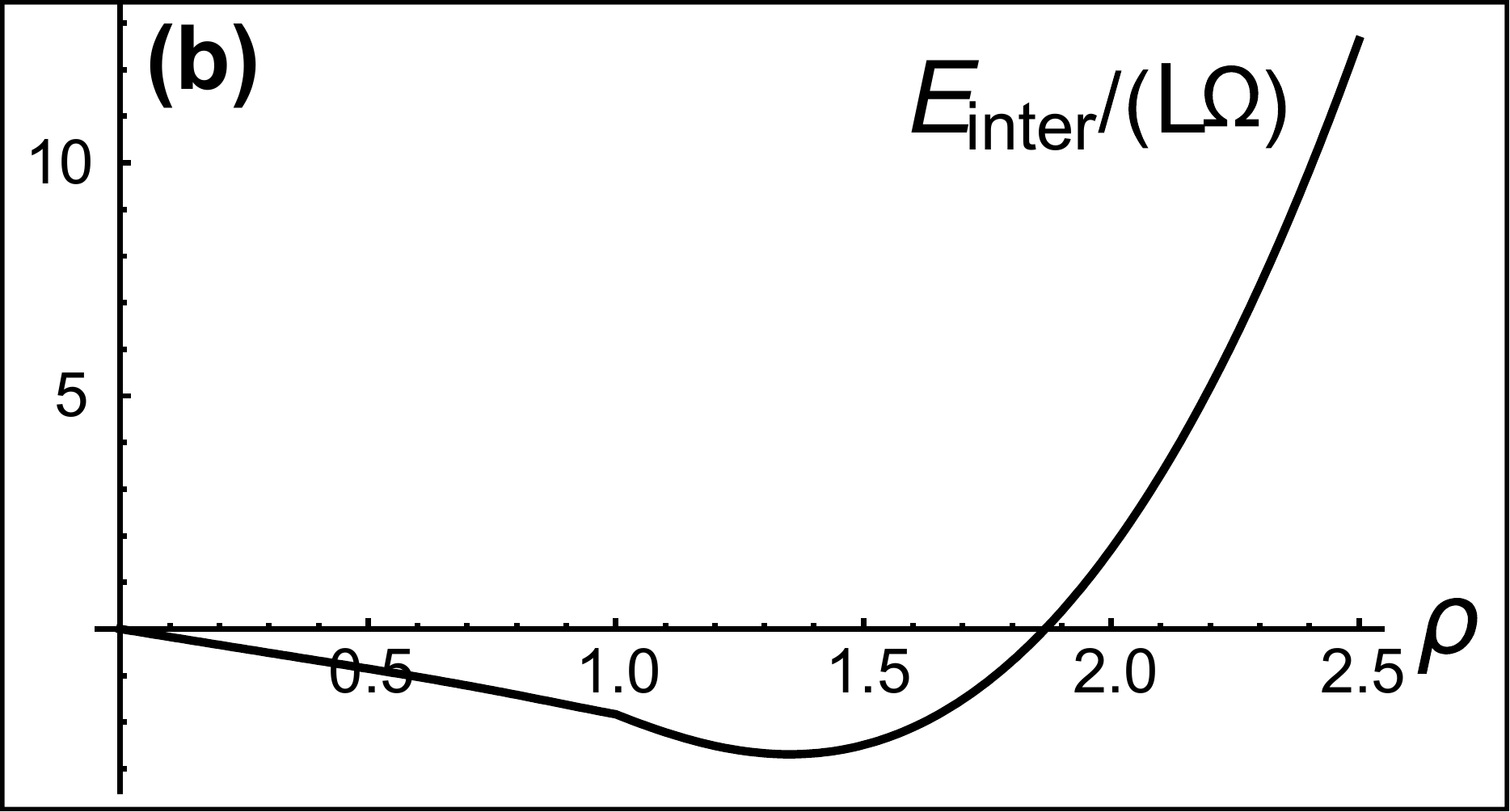}
\includegraphics[width=0.3\linewidth]{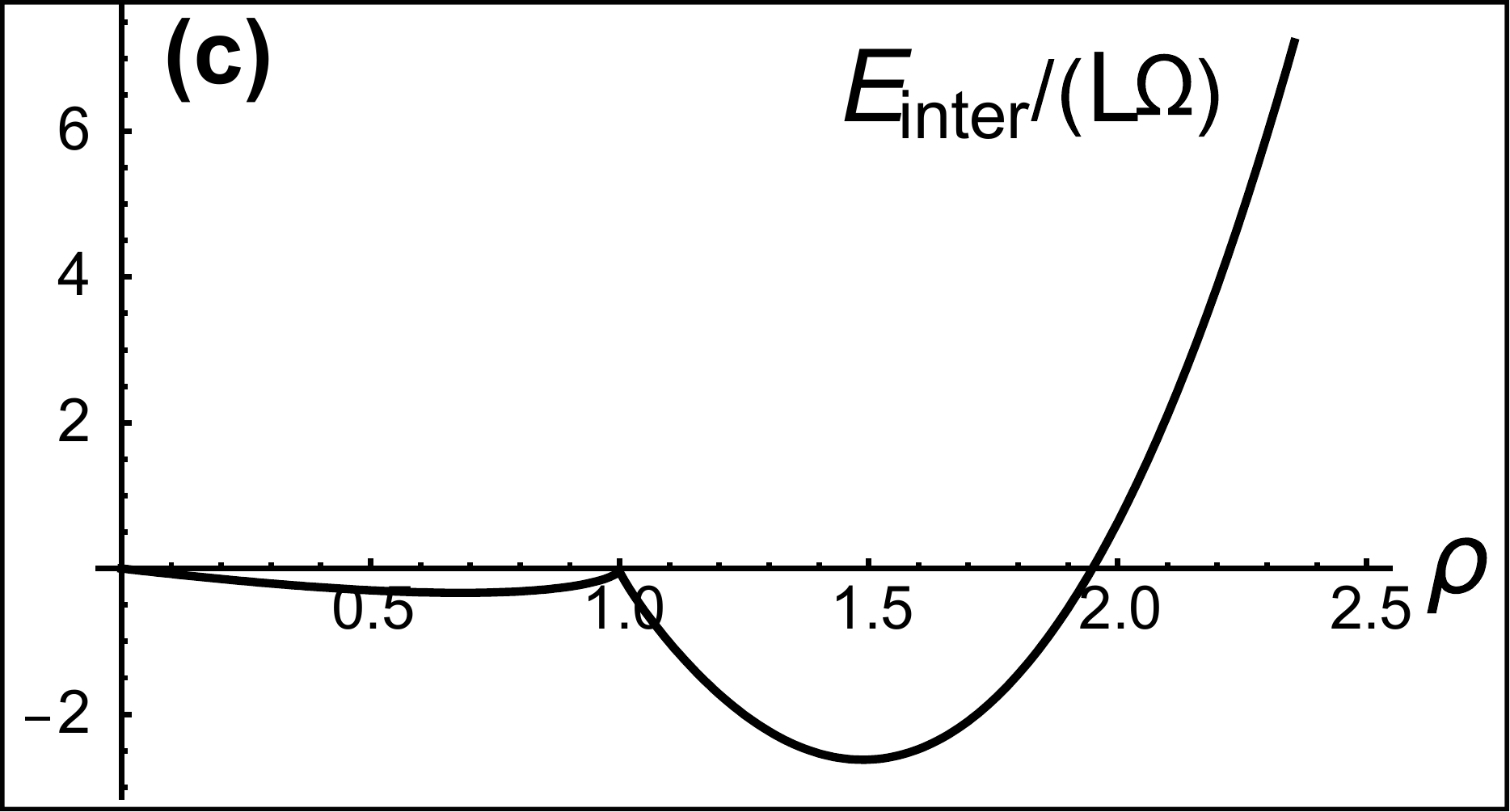}
\caption{
\label{Einter}
The dependence of interaction energy in leading order on pseudo-particle density $\rho$ at $\omega'=-1$ (a), $1$ (b), and $2$ (c); $\Omega'=0.3$.}
\end{figure}

(i) If $\omega < 2E_1$, all noninteracting pseudo-particles are bosons and their total energy is $E_{gr0}^{(nonint)}=M \omega$.

(ii) If $2E_1 < \omega < 2E_2$, two different scenarios are possible. The first case corresponds to the configuration with relatively small $M<(\omega - 2E_1)/2d$, such that all pseudo-particles are excited spins and their total energy is $E_{gr0}^{(nonint)}=2E_1M+M^2d$. The second case corresponds to the situation of relatively large $M>(\omega - 2E_1)/2d$, such that there are some bosons in the system. The number of spin excitations is $(\omega - 2E_1)/2d$ and their energy can be evaluated as explained above. The remaining $M-(\omega - 2E_1)/2d$ pseudo-particles are bosons each having an energy $\omega$. The total energy is a sum of contributions of spin and boson subsystems.

(iii) If $ \omega > 2E_2$, again two different possibilities have to be taken into account. In the first case, the pseudo-particle number is smaller than the total number of spins, $M<L$, so that there are no bosons. In the second case, it is larger than the total number of spins, $M>L$; hence there are $L$ spin excitations and $M-L$ bosons. It is straightforward to evaluate $E_{gr0}^{(nonint)}$ in both cases.

In Fig. (\ref{Einter}), we plot $E_{inter}/L\Omega$ as a function of $\rho$ at $\Omega'=0.3$ and three different values of the detuning $\omega'=-1$ (a), $1$ (b), and $2$ (c). In all cases, the hybridization between the spin and boson subsystems leads to the decrease of the total energy at small values of $\rho$. On the contrary, interaction increases the total energy at large values of $\rho$.

At large detunings, $E_{inter}$ as a function of $\rho$ has two cusps,
which can be attributed to distinct and poorly hybridized spin-like and boson-like states. At some critical detuning,
these two states become separated by a quantum phase transition at $\rho=1$ where $E_{inter}$ vanishes due to the vanishing of $\Delta'$.
It can be expected that finite-size corrections are of a particular importance in the vicinity of this point.

\subsection{Finite size corrections}

Leading-order finite size correction to the ground state energy beyond the mean-field approximation can be found from Eq. (\ref{sumh1final}). Positions of roots $x_l$ can be determined by using an approach of Ref. \cite{Altshuler}. The details of derivation are presented in Appendix E.

\begin{figure}[h]
\includegraphics[width=0.3\linewidth]{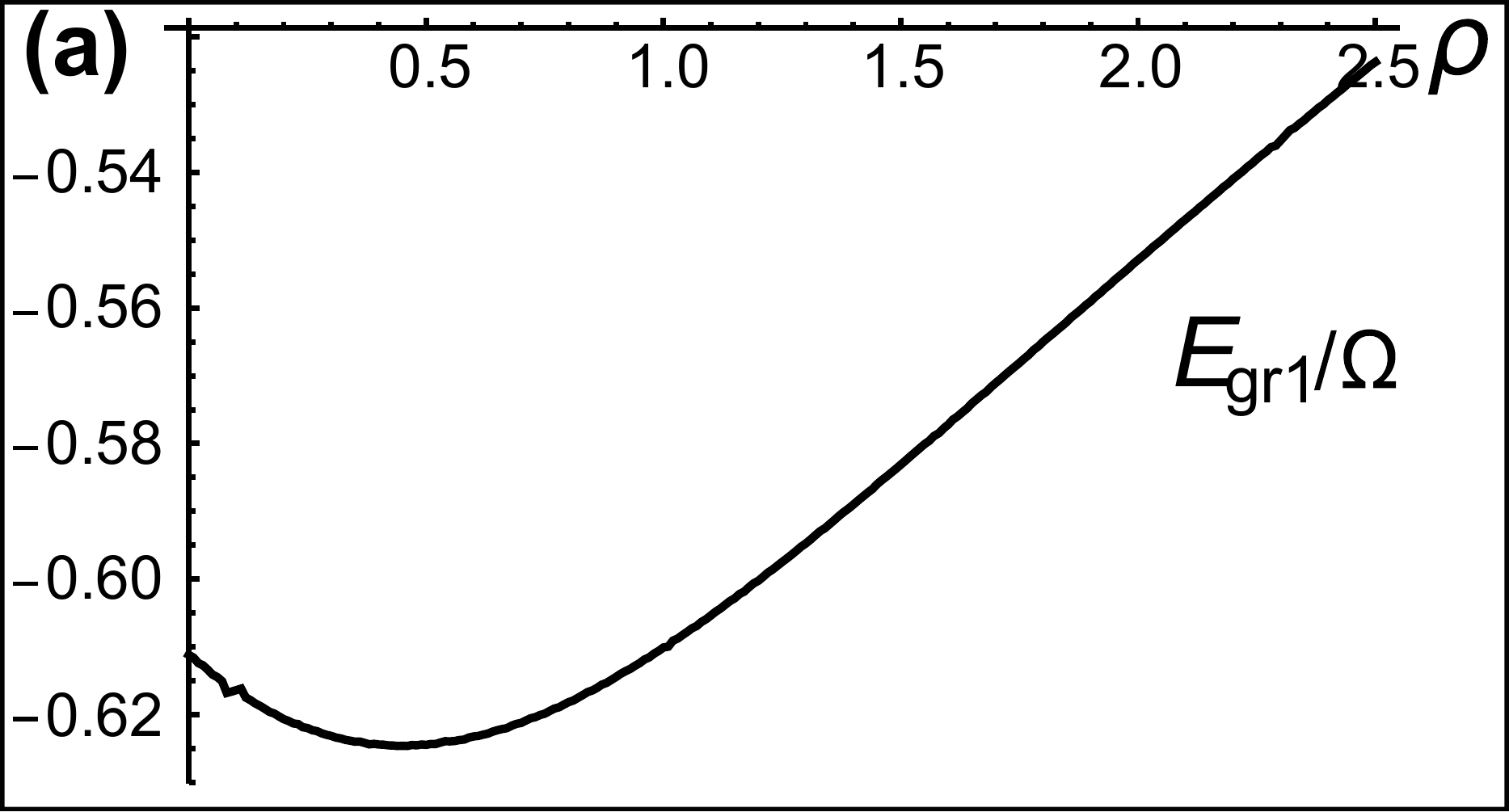}
\includegraphics[width=0.3\linewidth]{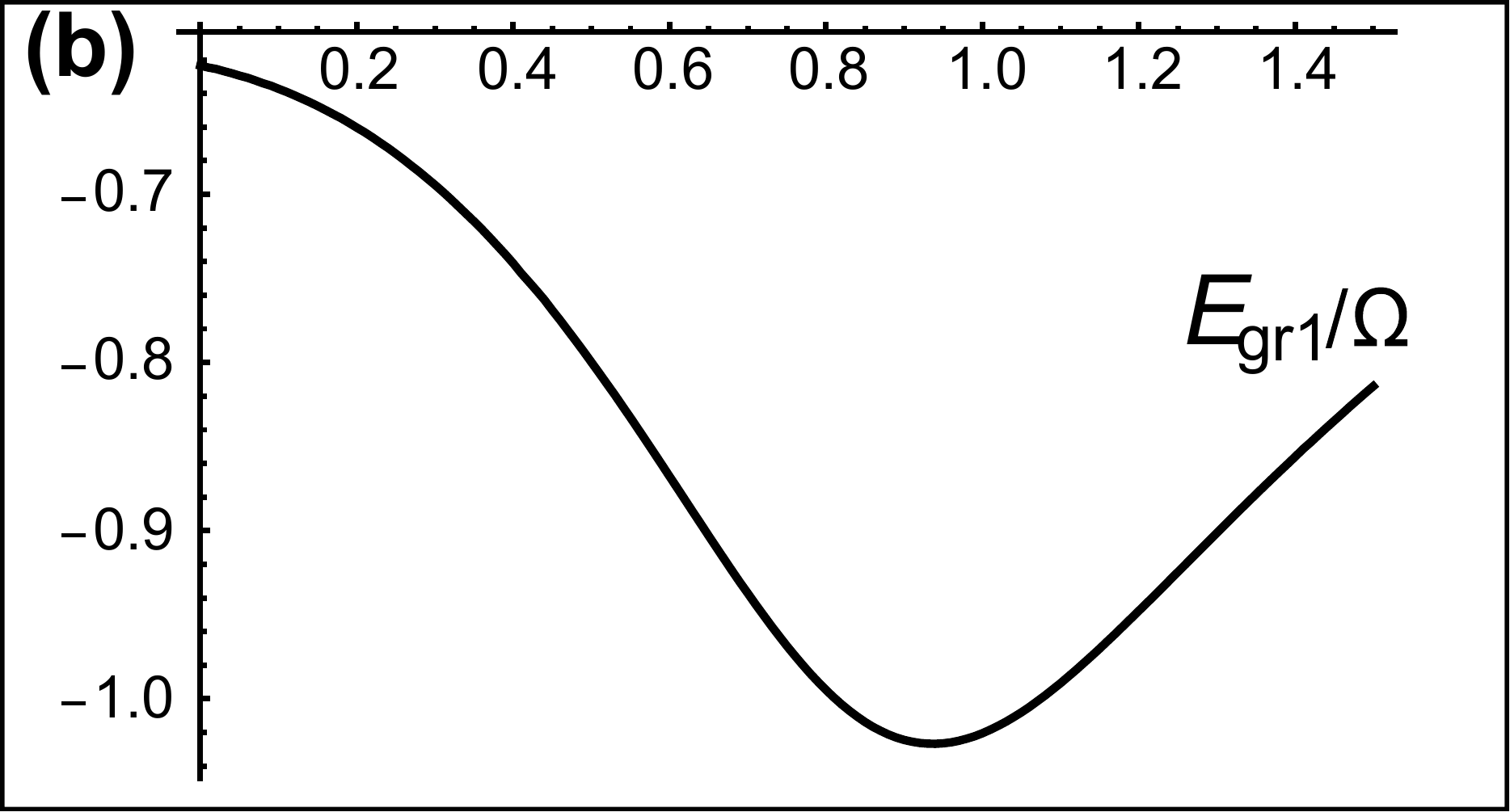}
\includegraphics[width=0.3\linewidth]{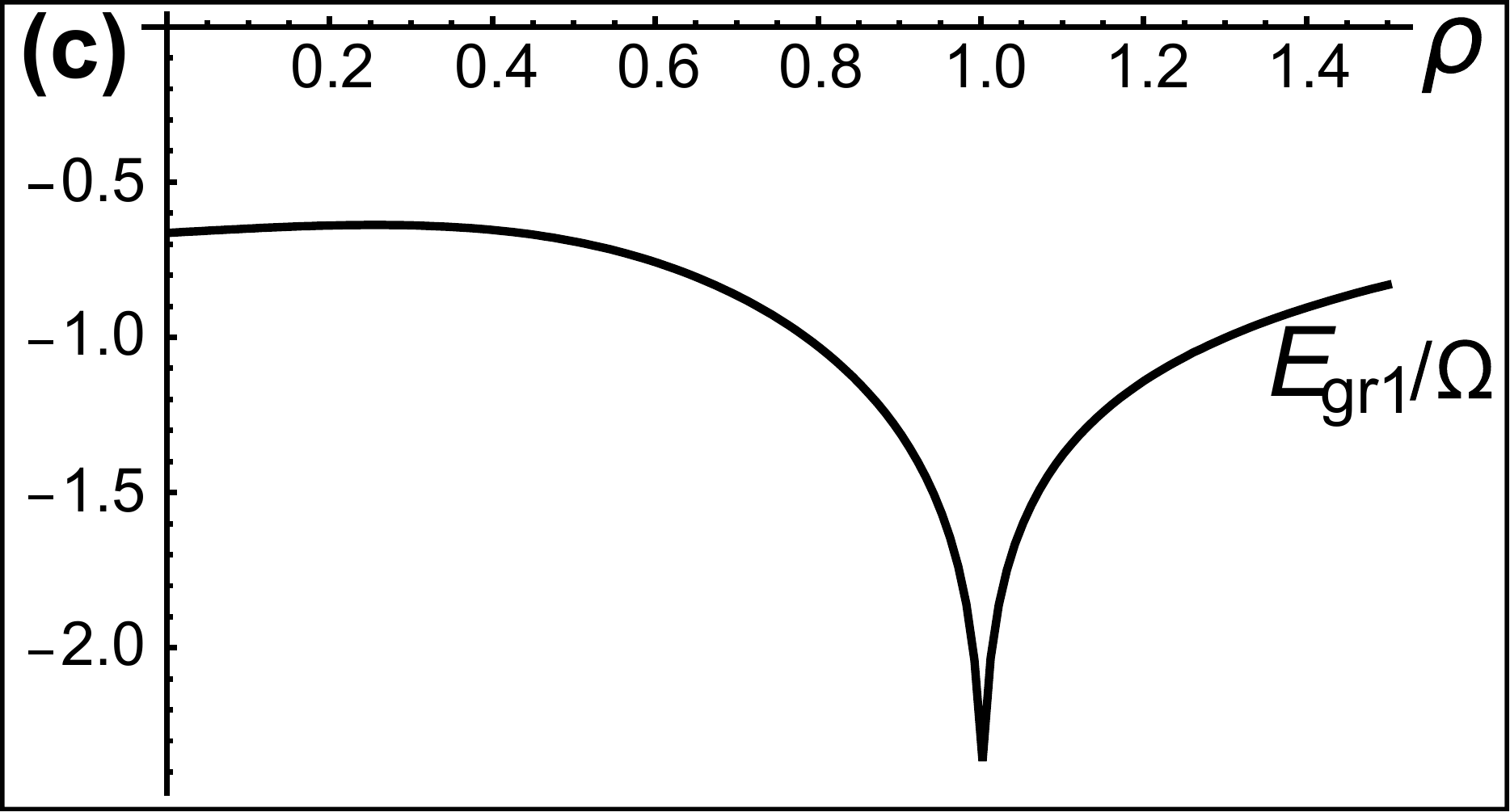}
\caption{
\label{gr1}
The dependence of the leading-order finite-size correction to the ground state energy (beyond the mean-field) on pseudo-particle density $\rho$ at $\omega'=-1$ (a), 1 (b), and 2 (c); $\Omega'=0.3$.}
\end{figure}

The results of our computations for $E_{gr1}$ are plotted in Fig. (\ref{gr1}) as a function of pseudo-particle density $\rho$ at $\Omega'=0.3$ and three different values of the detuning $\omega'=-1$ (a), 1 (b), and 2 (c). Note that, of course, we found that $E_{gr1} \sim E_{gr0}/L$.

A most important observation is that $-E_{gr1}$ is peaked in the vicinity of the point $\rho=1$ at large detuning, when $\Delta'$ as a function of $\rho$ starts to have a minimum of its first derivative. Surprisingly, peak appears even if $\Delta'(\rho=1)$ is nonzero, but $\Delta'(\rho)$ only shows a tendency of having a plateau. An interaction energy $E_{inter}$ goes to zero at $\rho=1$ at large detuning indicating a quantum phase transition. Consequently, finite-size corrections to the ground state energy become especially important in the vicinity of this phase transition and the corresponding region extends towards smaller detunings, for which the peak still exists. The fluctuative contribution $E_{gr1}$ lowers the total ground state energy. In particular, it tends to facilitate a realization of the superradiant transition and to shift the resulting optimal $\rho$ closer to $1$.

We now apply a similar approach to calculate finite-size corrections to the gap $ \Delta_1$ given by Eq. (\ref{E2diff}). The sums in (\ref{E2diff}) may again be evaluated by the method of Ref. \cite{Altshuler}.

\begin{figure}[h]
\includegraphics[width=0.3\linewidth]{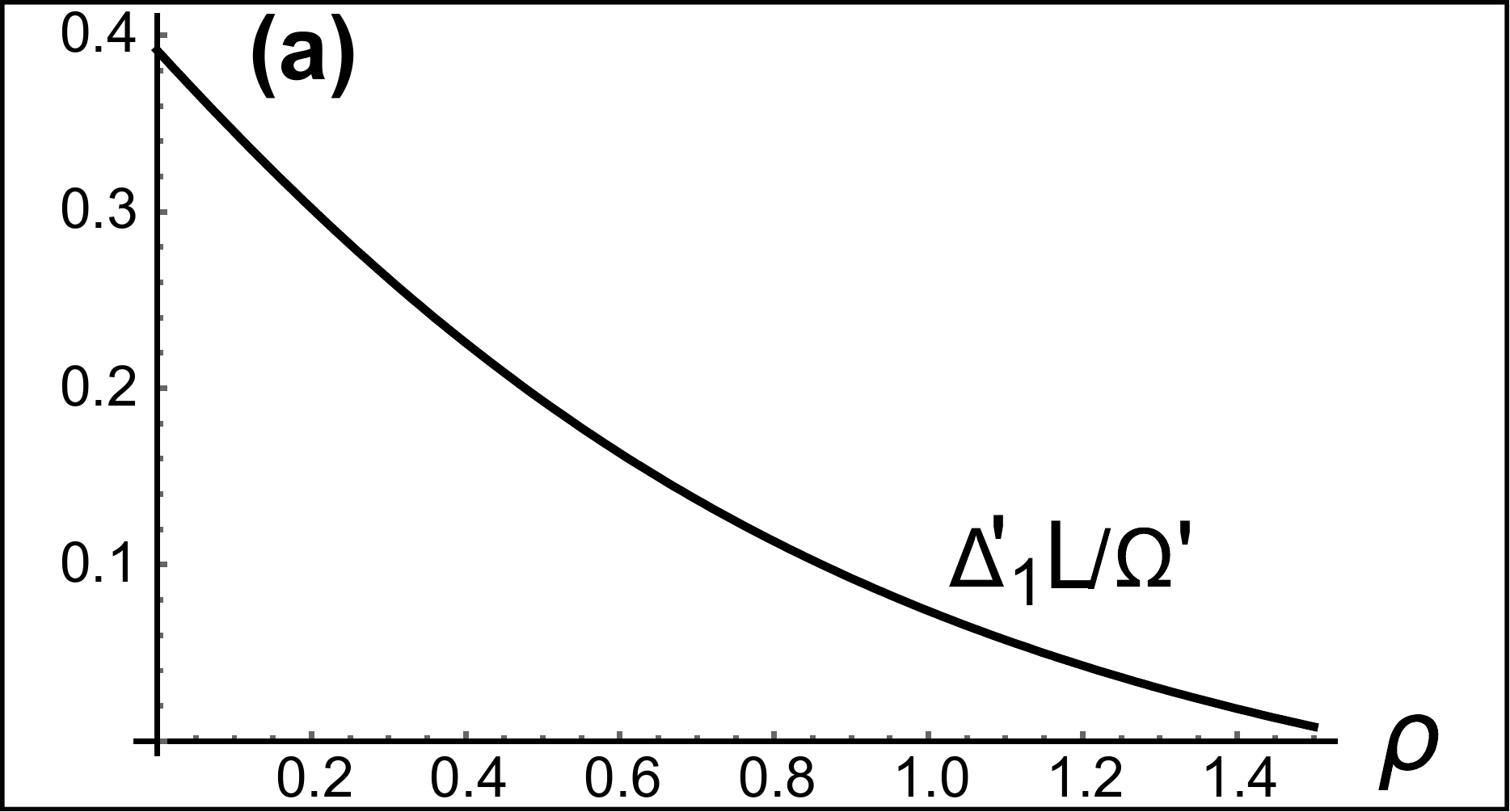}
\includegraphics[width=0.3\linewidth]{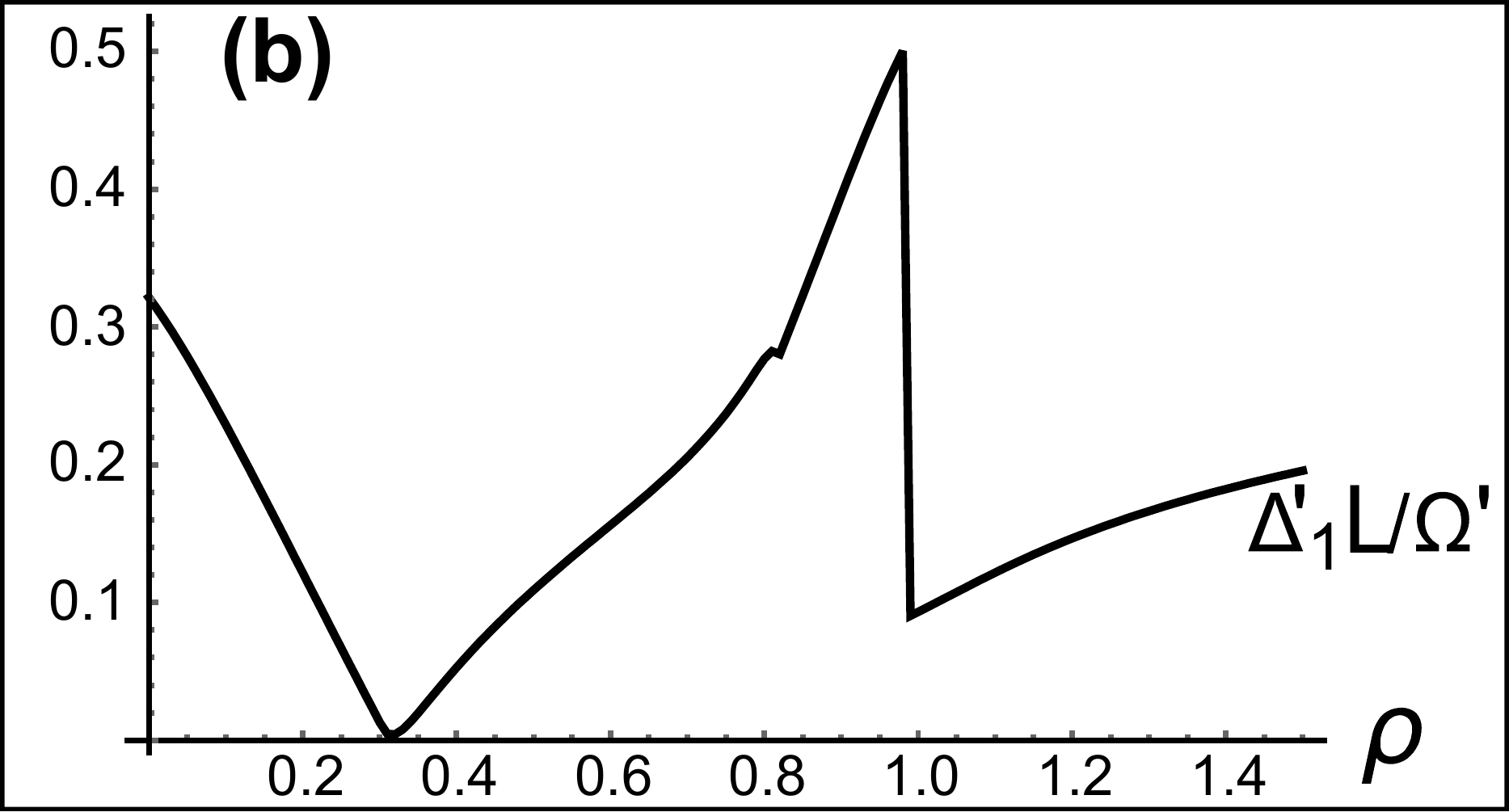}
\includegraphics[width=0.3\linewidth]{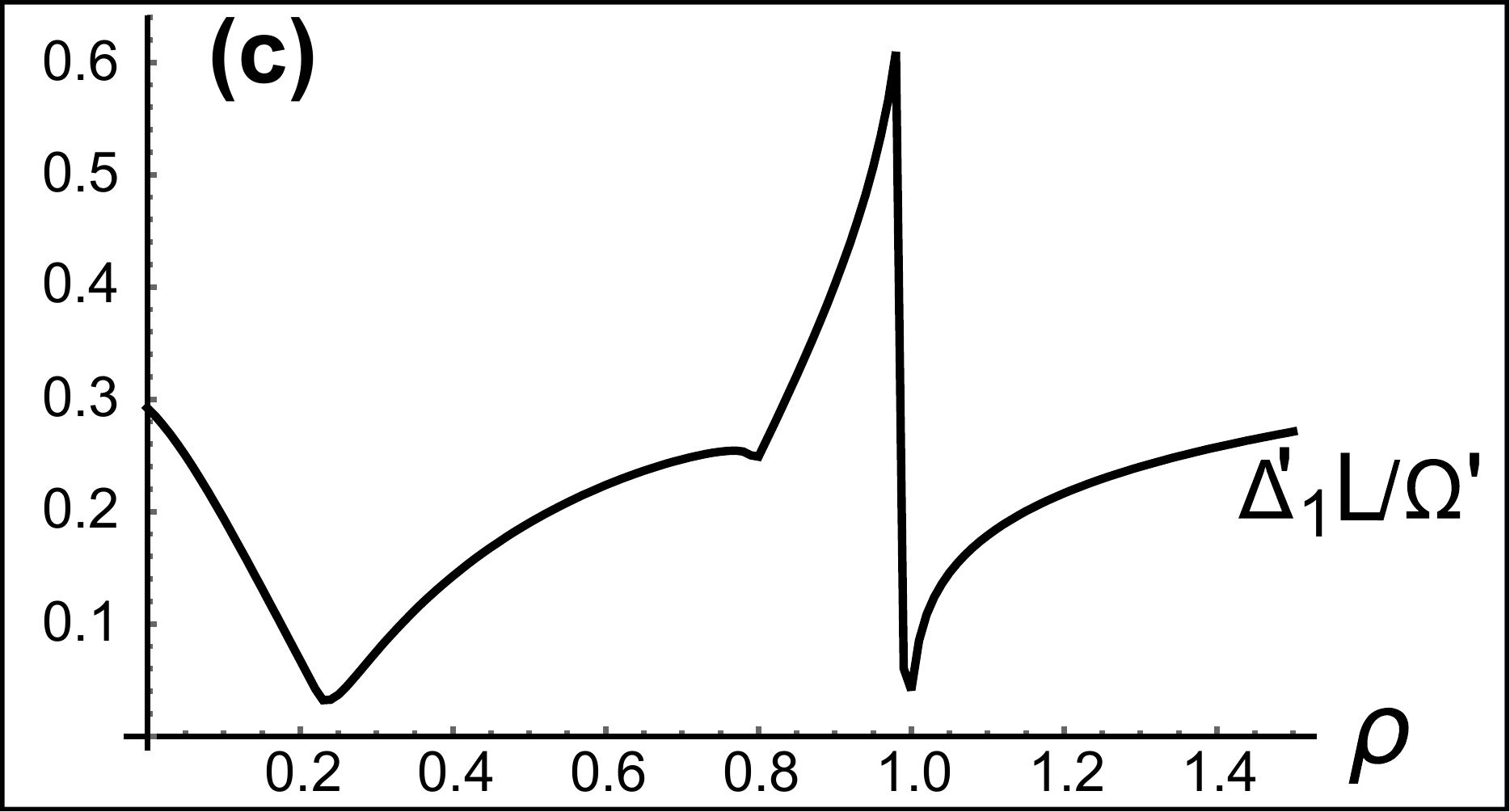}
\caption{
\label{gapcor}
The dependence of the leading-order finite-size correction to the spectral gap (beyond the mean-field) on pseudo-particle density $\rho$ at $\omega'=-1$ (a), 1 (b), and 2 (c); $\Omega'=0.3$.}
\end{figure}

The results of our computation for $ \Delta_1'= \Delta_1/g\sqrt{L}$ are plotted in Fig. \ref{gapcor} as a function of pseudo-particle density $\rho$ for $\Omega'=0.3$ and three different values of the detuning $\omega'=-1$ (a), 1 (b), and 2 (c). At zero and negative detuning, $ \Delta_1'$ is a monotonously decreasing function of $\rho$, as Fig. \ref{gapcor} (a) shows. In this case, $\Delta_{real}'$ is a monotonously increasing function. Therefore, finite-size corrections to $\Delta_{real}'$ are more important at small values of $\rho$, and their relative contribution smoothly decreases as $\rho$ grows. These corrections become even more significant at larger broadening $\Omega'$, when $\Delta_{real}'(\rho=0)=0$.

The situation, however, is different at large detunings, when $\Delta_{real}'$ has discontinuities of the first and second derivatives. In this regime, $ \Delta_1'$ also becomes a highly nontrivial function of $\rho$, shown in Fig. \ref{gapcor} (b) and (c). It is now characterized by peculiarities of the same kind as the features appearing in $\partial \Delta_{real}'$. Apparently, in addition to the vicinity of the point $\rho=0$, where a contribution of $ \Delta_1'$ is again of a particular importance, we see appearing a peak at $\rho=1$, where $\Delta_{real}'$ has a dip. We thus can conclude that a contribution of $ \Delta_1'$ to the total gap is especially significant around the point $\rho=1$, where it tends to smear out the local minimum in function $\Delta_{real}'$. The same conclusion was made for the finite-size correction to the ground state energy. Thus, the vicinity of this point is prone to quantum fluctuations even if order parameter $\Delta'$ does not vanish at $\rho=1$ but only has a minimum of its first derivative. This is a direct consequence of a quantum phase transition at $\rho=1$ and large detunings.

Let us stress that, of course, quantum fluctuations are also able to smear out two transitions corresponding to the discontinuities of second derivatives of $\Delta_{real}'$.

There also exist finite-size corrections originating from the replacement of sums by integrals in the equations for the order parameter (\ref{gap}) and chemical potential (\ref{gapchemeqs}), as well as in the expression of the ground state energy (\ref{energygr0}). These corrections can be found by performing a more accurate replacement, as it was done, for example, in Ref. \cite{Altshuler} for Richardson model. We are not going to present a detailed analysis here. Instead we explain our main findings on a qualitative level. We found that such corrections both to the ground state energy and to the gap are also most significant in the vicinity of the point $\rho=1$ at large detuning, as well as at $\rho=0$.

\section{Conclusions}

Dicke model, which is well known from quantum optics, is directly related to the family of Richardson-Gaudin models and may be viewed as their extension, since they represent certain limiting cases of Dicke model. Bethe equations for Richardson-Gaudin models in the thermodynamical limit can be solved explicitly using an approach developed by Gaudin \cite{Gaudin} and Richardson \cite{Rich3} decades ago. Moreover, this approach allows for an iterative evaluation of finite-size corrections both to the ground state and lowest excited state energies. Such corrections are of importance for the crossover region from the few-'particle' limit of the system to the macroscopic regime.

In the present article, we extended the method of Richardson \cite{Rich3} to Bethe equations for the Dicke model. Namely, we presented formal expressions for the low-lying part of the energy spectrum as well as leading order finite-size corrections for a completely generic distribution of individual spin energies. These results can be applied for the crossover from the thermodynamical limit, which is correctly described by the mean-field approximation, to the fluctuation-dominated regime in small systems. They also provide an additional link between the Dicke model and the Richardson-Gaudin family of models.

We then applied our results for the simplest case of equally-spaced distributions of individual spin energies over some interval of finite width. We found a quite reach zero-temperature phase diagram and studied in some details 'fine structure' of the gap in the excitation spectrum, which can experience discontinuities of its derivatives as a function of a pseudo-particle density. We also analyzed various contributions to the ground state energy, which are responsible, for example, for the superradiant transition. Finally, we determined regions on the phase diagram, where quantum fluctuations are of particular importance for both the ground state and the low-energy excited states.

\begin{acknowledgments}

Useful comments by V. I. Yudson, A. A. Elistratov, and S. V. Remizov are acknowledged. This work was supported by RFBR (projects nos. 15-02-02128 and 14-02-00494), by Ministry of Education and Science of Russia (grants nos. 14.Y26.31.0007 and 02.A03.21.0003 from 27.08.2013), and by Russian Science Foundation (contract no. 16-12-00095). D. S. S. acknowledges a support by the Fellowship of the President of Russian Federation for young scientists (fellowship no. SP-2044.2016.5).
\end{acknowledgments}

\appendix

\section{Two expansions of field $F(z)$}

Let us represent $F(z)$ as a multipole expansion
\begin{eqnarray}
F(z)=\sum_{s=-1}^{\infty}F^{(s)}z^{-s}.
\label{multi}
\end{eqnarray}
By substituting this expansion into (\ref{Rikatya}), we find
\begin{eqnarray}
& F^{(-1)}=2/g^2 , \label{momentminus1}\\&
F^{(0)}=- \omega/g^2 , \label{momentzero} \\&
F^{(1)}=(2M-L)/2 , \label{moment1} \\&
F^{(2)}=-\frac{g^2}{4} \sum_{n} H(n) - \frac{1}{2} \sum_{n} \epsilon_n + \frac{M\omega}{2}.
\label{moments}
\end{eqnarray}
Compared to the Richardson model, a nonzero $F^{(-1)}$ appears in the multipole expansion, which modifies a whole derivation. The first three relations (\ref{momentminus1})-(\ref{moment1}) satisfy Eq. (\ref{Fz}) automatically. Using (\ref{Fz}), we also find that $F^{(2)}$ can be represented as
\begin{eqnarray}
F^{(2)}=\sum_{j} e(j) - \frac{1}{2} \sum_{n} \epsilon_n.
\label{F2}
\end{eqnarray}
By comparing (\ref{moments}) and (\ref{F2}), we find
\begin{eqnarray}
\sum_{j} e(j)= - \frac{g^2}{4} \sum_{n} H (n) + \frac{M\omega}{2}.
\label{EjF2}
\end{eqnarray}

We also expand $F(z)$, $H(n)$, and $E$ in powers of $1/L$ as $F(z)=\sum_{k=0}^{\infty}F_k(z)$, where $F_0 \sim L$. For the first three terms, we obtain from the field equation
\begin{eqnarray}
F_{0}^{2} = \frac{4M}{g^2}
+\left( \frac{1}{2}\sum_{n} \frac{1}{z - \epsilon_n}+\frac{\omega}{g^2}-\frac{2z}{g^2}\right)^2-
\sum_{n} \frac{H_{0}(n)}{z - \epsilon_n}.
\label{RikatyaF0}
\end{eqnarray}
\begin{eqnarray}
2F_0 F_1 = \frac{1}{2}\sum_{n} \frac{1}{(z - \epsilon_n)^2}-
\sum_{n} \frac{H_1(n)}{z - \epsilon_n} - \frac{dF_0}{dz}+\frac{2}{g^2}.
\label{RikatyaF1}
\end{eqnarray}
\begin{eqnarray}
2F_0 F_2 = - \sum_{n} \frac{H_2(n)}{z - \epsilon_n} - F_1 ^2 - \frac{dF_1}{dz},
\label{Fsecondorder}
\end{eqnarray}
where
\begin{eqnarray}
H_k(n)=-\frac{1}{2\pi i }\oint_{C_{e}}  \frac{F_k(z)dz}{z - \epsilon_n}.
\label{Hnint}
\end{eqnarray}

Note that (\ref{RikatyaF0}) and (\ref{RikatyaF1}) are different from corresponding equations in Richardson model, while (\ref{Fsecondorder}) is the same.

\section{Contour-integral representation of $H(n)$}

In order to derive Eqs. (\ref{ksai})-(\ref{gapchemeqs}), following relation is utilized
\begin{eqnarray}
H(n)=-\frac{1}{2\pi i } \left(  \oint_{C_{e,\epsilon}}  \frac{F(z)dz}{z - \epsilon_n}
- \oint_{C_{\epsilon}}  \frac{F(z)dz}{z - \epsilon_n}\right),
\label{Hnrepl}
\end{eqnarray}
where a closed contour $C_{e,\epsilon}$ clockwises around all the singularities of $F$ coming from poles at all roots $e(j)$ as well as all energies $\epsilon$, while $C_{\epsilon}$ includes only all energies $\epsilon$. For the first integral in the right-hand side of the above equation we use the expression of $F(z)$ given by (\ref{Fz}). We then find
\begin{eqnarray}
H(n)=-\frac{2\epsilon_n}{g^2}+\frac{\omega}{g^2}
+\frac{1}{2\pi i } \oint_{C_{\epsilon}}  \frac{F(z)dz}{z - \epsilon_n},
\label{Hnfin}
\end{eqnarray}
where for the last integral in a leading order we may use the ansatz (\ref{F0ansatz}).

\section{Finite-size corrections to the ground state energy}

Let us now consider Eq. (\ref{RikatyaF1}) for $F_1$. The set of quantities $H_1(n)$ can be determined by requirement that $F_1$ does not introduce any new charges located on real axis, so that $F_0$ describes all such charges. As follows from Eq. (\ref{RikatyaF1}), this is possible only if the right-hand side vanishes at real zeros $x_l$ of $F_0$, otherwise $F_1$ has poles at these zeros. In contrast to the Richardson model, the number of zeros $x_l$ is equal to the number of energy levels, i.e., to $L$, as it is clear from Eq. (\ref{F0ansatz}).

Thus, we require that the right-hand side of equation (\ref{RikatyaF1}) vanishes at $z=x_l$. By substituting (\ref{F0ansatz}) to (\ref{RikatyaF1}) at $z=x_l$ and representing $H_1(n)$ as
\begin{eqnarray}
H_1(n)=\frac{1}{\eta _n} \sum_{l} \frac{h_1(l)}{x_l-\epsilon_n},
\label{H1ndecomp}
\end{eqnarray}
we find
\begin{eqnarray}
h_1(l)=\frac{1}{2}\left(\frac{N_l}{D_l} - \zeta_l  \right),
\label{h1fin}
\end{eqnarray}
where
\begin{eqnarray}
\zeta_l = \sqrt{(x_l-a)(x_l-a^{*})},  \notag \\
N_l=\frac{4}{g^2}+\sum_{n}\frac{1}{(x_l-\epsilon_n)^2},  \notag \\
D_l=\sum_{n}\frac{1}{\eta_n}  \frac{1}{(x_l-\epsilon_n)^2}.
\label{h1defin}
\end{eqnarray}
Note that the expression of $N_l$ is different from that for Richardson model.

In order to find a first-order correction to the energy, we need to obtain $ \frac{g^2}{4} \sum_{l}H_1(l)$ which reduces to $\sum_{l}h_1(l)$ due to Eq. (\ref{H1ndecomp}). This sum can be evaluated by using Eq. (\ref{h1fin}) and by simplifying the ratio of sums (see., e.g., Ref. \cite{Altshuler}). Finally, we arrive at Eq. (\ref{sumh1final}).

A similar approach can be used to simplify the expression of $F_1$ using Eqs. (\ref{RikatyaF1}) and (\ref{h1fin}). After some algebra, we obtain
\begin{eqnarray}
F_1 = \frac{1}{2Z} \left(\sum_{n} \frac{z+\epsilon_n-2 \lambda}{Z+ \eta_n}
- \sum_{l} \frac{z+x_l-2 \lambda}{Z+ \zeta_l} - \frac{z-\lambda}{Z} +1 \right).
\label{F1simpl}
\end{eqnarray}
Let us mention that $F_1$ has no monopole moment in contrast to the similar result for Richardson model.

\section{Excited states}

We now consider a configuration with one isolated free charge located out of the line of charges. Such an electrostatic configuration corresponds to the excited state of Dicke model. We denote the total field due to all charges as $F'(z)$. It satisfies Eq. (\ref{Rikatya}). The field due to the isolated charge located at $e(1)$ is $1/(z-e(1))$. Let us expand $e(1)$ in power series in $1/L$ as $e(1)=e_1+e_2$. Then, we split $F'$ into the contribution from this isolated charge and remaining charges $\overline{F'}$
\begin{eqnarray}
F'=\overline{F'}+\frac{1}{z-e(1)}.
\label{split1}
\end{eqnarray}
We expect that the substraction of a single charge from the whole system of charges does not change the total field in the leading order. Hence, we can write
\begin{eqnarray}
\overline{F'}=F_0+\overline{F'_1}.
\label{split2}
\end{eqnarray}
Thus, the total first-order correction to $F_0$ is $\frac{1}{z-e(1)}+\overline{F'_1}$. It must satisfy Eq. (\ref{RikatyaF1}). Let us represent it as
\begin{eqnarray}
2F_0 \overline{F'_1} = - \frac{dF_0}{dz}+\frac{2}{g^2}+\frac{1}{2}\sum_{n} \frac{1}{(z - \epsilon_n)^2}-
\sum_{n} \frac{H' _1(n)}{z - \epsilon_n} - 2F_0 \frac{1}{z-e_1}.
\label{RikatyaF1prime}
\end{eqnarray}
The function $\overline{F'_1}$ should not have poles on real axis. Therefore, $e_1$ has to coincide with one of the zeros $x_l$ of $F_0$. Note that has a term $2/g^2$ absent in the case of Richardson model.

If $z=x_m$ and $m \neq l$, we have from (\ref{RikatyaF1prime}) the same equation as previously. However, for $z=x_l$ we have a modified equation
\begin{eqnarray}
\sum_{n} \frac{H'_1(n)}{x_l - \epsilon_n} = - 3 \frac{dF_0}{dz}(z=x_l)+\frac{2}{g^2}+\frac{1}{2}\sum_{n} \frac{1}{(x_l - \epsilon_n)^2},
\label{H1l}
\end{eqnarray}
where we used an identity
\begin{eqnarray}
\frac{F_0 (z)}{z-x_l} (z\rightarrow x_l)= \frac{dF_0}{dz}(z=x_l)
\end{eqnarray}
Simple algebra shows that a contribution to the total energy due to this quantity is $2\zeta_l=2\sqrt{(x_l-\lambda)^2 + \Delta^2}$.

A correction to the excited state energy is the same as in Richardson model \cite{Rich3}
\begin{eqnarray}
 \Delta_1= \frac{1}{D_l} \left( (\overline{F'_1}^2 - F_1 ^2) + \frac{d}{dz}(3 \overline{F'_1} - F_1)+e_2\frac{d^2 F_0}{d z^2} \right)_{z=x_l} + \sum_{m \neq l} \frac{1}{D_m} \left( (\overline{F'_1}^2 - F_1 ^2) + \frac{d}{dz}(\overline{F'_1} - F_1)+\frac{2 \overline{F'_1}}{z-x_l} \right)_{z=x_m},
\label{E2diff}
\end{eqnarray}
where $D_l$ is given by Eq.  (\ref{h1defin}), $e_2$ is a correction to the position of the isolated charge given by
\begin{eqnarray}
e_2=\frac{1}{\zeta_l D_l}\left( \frac{x_l-\lambda}{\zeta_l^2} - F_1 (x_l)\right),
\label{}
\end{eqnarray}
while
\begin{eqnarray}
\overline{F'_1} = F_1+ \frac{1}{z - x_l} \left(\frac{\zeta_l}{Z}-1\right),
\label{F1overline}
\end{eqnarray}
and $F_1$ is a first-order correction to the field, given by (\ref{F1simpl}).

\section{Real roots of $F_0(z)$ for the equally-spaced model}

All roots $x_l$ of $F_0$ except of $x_{max}$ are confined between two neighboring spin energies. We, therefore, can represent them as $x_l=\epsilon_l+\delta_l$, where $\delta_l < d$. Keeping leading order in $\delta_l$, we obtain
\begin{eqnarray}
E_{gr1}=
\sum_{l=0}^{L-2} \frac{\delta_l (\epsilon_l-\lambda)}{\sqrt{(\epsilon_l-\lambda)^2+\Delta^2}}+\frac{1}{2} (2\lambda-\omega)
+
\sqrt{(x_{max}-\lambda)^2+\Delta^2}-\sqrt{(E_{2}-\lambda)^2+\Delta^2},
\label{Eground11}
\end{eqnarray}
while $x_l$ are determined by the equation
\begin{eqnarray}
\sum_{n=0}^{L-1}\frac{1}{x_l-\epsilon_n}\frac{1}{\sqrt{(x_l-\epsilon_n)^2+\Delta^2}}=\frac{4}{g^2}.
\label{xleq}
\end{eqnarray}
Following Ref. \cite{Altshuler}, we represent the left-hand side of this equation as a sum of two contributions. The first one is a discrete sum over energy levels, which are not too far from $x_l$, whereas the second one is a principal value integral over remaining levels. Solving the resulting equation for $\delta_l$, we obtain
\begin{eqnarray}
\frac{\pi\delta_l}{d}=\cot^{-1} \frac{a(\epsilon)}{\pi},
\label{arccot}
\end{eqnarray}
where
\begin{eqnarray}
a(\epsilon)=\log \frac{\epsilon-E_1}{E_2-\epsilon} \frac{\sqrt{(\epsilon-\lambda)^2+\Delta^2}\sqrt{(E_2-\lambda)^2+\Delta^2}+(\epsilon-\lambda)^2+\Delta^2 + (\epsilon-\lambda)(E_2-\epsilon)}
{\sqrt{(\epsilon-\lambda)^2+\Delta^2}\sqrt{(E_1-\lambda)^2+\Delta^2}+(\epsilon-\lambda)^2+\Delta^2 + (\epsilon-\lambda)(E_1-\epsilon)}-\frac{4\Omega}{g^2L}\sqrt{(\epsilon-\lambda)^2+\Delta^2},
\label{aepsilon}
\end{eqnarray}
and $\epsilon=E_1+ld$.

There is an additional solution $x_{max} > E_2$, which does not exist in the case of Richardson model.  In the thermodynamical limit, it can be found by switching from summation to the integration in (\ref{F0ansatz}) and then solving the equation $F_0(z)=0$. It is convenient to introduce a dimensionless variable $x_{max}'$ defined as $x_{max}'=(x_{max}-\lambda)/g\sqrt{L}$. We readily find that it satisfies the transcendental equation
\begin{eqnarray}
& 4 \Omega' \sqrt{(x_{max}'-\lambda')^2+\Delta'^2} = \nonumber \\
& \log \left(  \frac{\Omega'/2+x_{max}'}{-\Omega'/2+x_{max}'} \frac{\sqrt{(\Omega'/2-\lambda')^2+\Delta'^2}\sqrt{(x_{max}'-\lambda')^2+\Delta'^2}+(x_{max}'-\lambda')^2+\Delta'^2 - (x_{max}'-\lambda')(x_{max}'-\Omega'/2)}
{\sqrt{(\Omega'/2+\lambda')^2+\Delta'^2}\sqrt{(x_{max}'-\lambda')^2+\Delta'^2}+(x_{max}'-\lambda')^2+\Delta'^2 - (x_{max}'-\lambda')(x_{max}'+\Omega'/2)}\right),
\label{xmaxeq}
\end{eqnarray}
which can be solved numerically.

We can substitute $\delta_l$ and $x_{max}$ to the expression for the correction to the ground state energy (\ref{Eground11}) and then switch from summation over $l$ to the integration over $\epsilon$. The integral can be readily evaluated numerically.

\end{document}